\documentclass[10pt,journal,compsoc]{IEEEtran}

\ifCLASSOPTIONcompsoc
   \usepackage[nocompress]{cite}
\else
    \usepackage{cite}
\fi

\ifCLASSINFOpdf
 
\else
 
\fi

\usepackage{graphicx}
\usepackage{xcolor}
\usepackage{tabularx}

\usepackage{authblk}

\begin{document}

\title{Turning the Hunted into the Hunter via Threat Hunting: Life Cycle, Ecosystem, Challenges and the Great Promise of AI}

\author{Caroline Hillier}
\author{Talieh Karroubi}
\affil{School of Computer Science, University of Guelph, ON, Canada, \\chilli04@uoguelph.ca, tkarroub@uoguelph.ca}




\IEEEtitleabstractindextext{%
\begin{abstract}
\textcolor{black}{The threat hunting lifecycle is a complex atmosphere that requires special attention from professionals to maintain security. This paper is a collection of recent work that gives a holistic view of the threat hunting ecosystem, identifies challenges, and discusses the future with the integration of artificial intelligence (AI). We specifically establish  a life cycle and ecosystem for privacy-threat hunting in addition to identifying the related challenges. We also discovered how critical the use of AI is in threat hunting. This work paves the way for future work in this area as it provides the foundational knowledge to make meaningful advancements for threat hunting.  }
\end{abstract}

}

\maketitle

\IEEEdisplaynontitleabstractindextext

\IEEEpeerreviewmaketitle

\section{Introduction}

The process of threat hunting is involved with proactive use of manual or machine-based methods by cybersecurity analyst to find security incidents or threats that previously spread automatic detection techniques missed. In order for analysts to succeed at threat hunting, they need to understand how to arrange their tools into detecting the threats\cite{Pro-Def-Jour001}. In addition to having sufficient knowledge of malware, exploits and network protocols, they need to be able to navigate the vast quantities of data, including logs, metadata, and packet capture (PCAP) data. Proactive defence has received a research focus in recent years \cite{BUT-Chap001}\cite{ZXMN-Chap001}.

There are systems like Intrusion Detection Systems (IDSs)\cite{a1} that have reactive mechanism in nature and detects intrusions which have already been in the system. As a result, reactive mechanisms are far behind and are not be able to handle actions taken by clever adversaries. However, there is proactive defense that has been designed to detect potential attackers and/or mitigate the impact of intrusions ahead of their penetration like IPSs \cite{ZXMN-Chap001}.

\begin{figure}
    \centering
    \includegraphics[width=8cm]{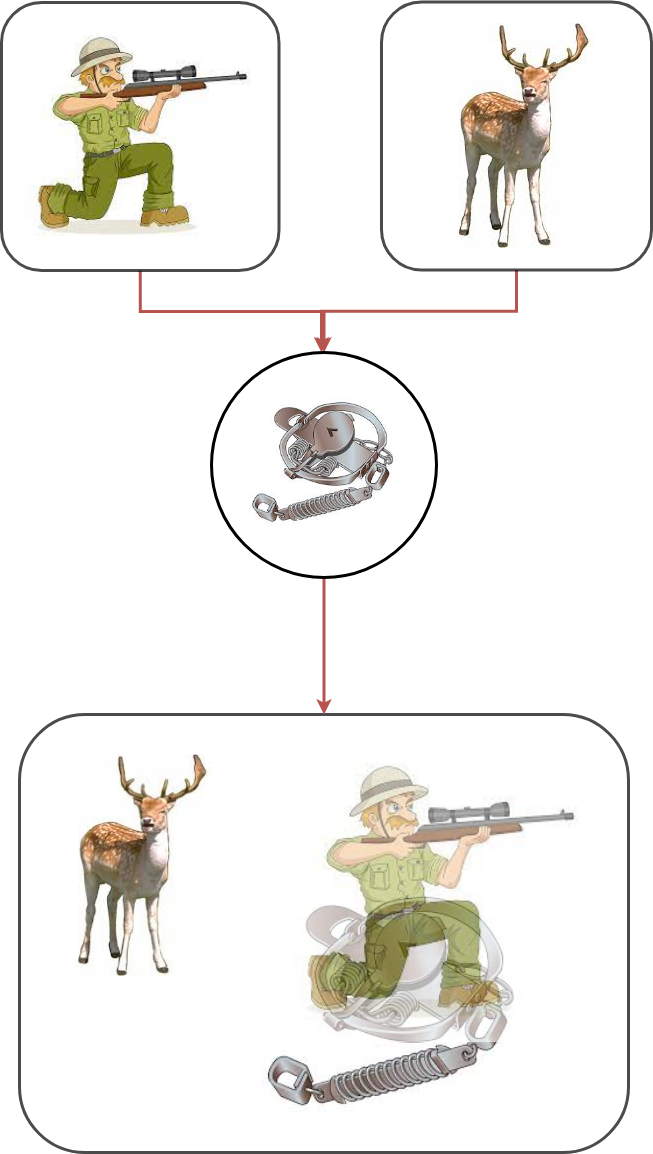}
    \caption{Threat Hunting: Turning the Hunted into the Hunter}
    \label{fig:sec2}
\end{figure}

Threat hunting is a branch of proactive defense that deals with emerging and unseen threats. It is the act of detecting and eliminating cyberattacks that have penetrated your environment without creating any alarms, unlike traditional cybersecurity investigations and responses, which activated by system alerts, and are carried out after possibly malicious activity has been spotted. Therefore, a threat hunter is discovering and uncovering new threats\cite{CISCO-URL001}. 
\textcolor{black}{Figure \ref{fig:sec2} shows how threat hunting turns the hunter into the hunted. The malicious actor that is posing the threat becomes the target of the hunt by cybersecurity professionals. }  \textcolor{black}{Indeed, in Figure \ref{fig:sec2} the hunter represents the malicious actor that is trying to hack into a target entity. The deer represents the entity that is the malicious actor's target. }

Threat hunting is a sensitive area which requires attention \cite{a2}. Though the cybersecurity tools developed by companies are strong, adversaries frequently find a way invade the system. Therefore threat hunters have the responsibility to find adversaries quickly, to prevent further damage. 
Within threat hunting there is a wide array of stages and areas of applications so this broad topic is challenging to fully understand  \cite{a3}. This literature review survey aims to provide a holistic understanding of current work regarding threat hunting. Although there are some surveys regarding threat hunting, there is insufficient work to fully investigate current advancements and their benefits as well as upcoming need related to this area. In this paper we discuss the recent developments and their relations with each other and conclude the possible future predictions of the integration of AI and other future applications.

Big industrial companies such as CISCO \cite{CISCO-URL001}\cite{CISCO-URL002}, Palo Alto \cite{Palo-Alto-URL001}  and Dragos \cite{Dragos-URL001} are working on threat hunting. In recent years, threat hunting has been of great interest to the research community as well \cite{YRTIE-Chap001}\cite{YRTIE-Chap002}. Academia is following the industry via defining theses \cite{Master-Thes001}\cite{Master-Thes002} and projects \cite{Proj-URL001}, offering courses \cite{Course-URL001},
	building labs \cite{Lab-Jour001}\cite{Lab-Conf002}, presenting lectures \cite{Lect-URL001} and organizing competitions \cite{Threat-Hunt-Conf010}. However, the academic research community still looks lagging behind the industry. \textcolor{black}{Although there might be some related survey papers, they have shortcomings, which motivate our work in this survey. Some of them are very general others are too specific.}

The rest of this paper is organized as follows. Section \ref{ExSurv} presents a review on existing survey and their shortcomings in order to highlight our motivations for our work in this paper. Section \ref{Emerg} discusses the problem of emerging threats. Sections \ref{Life}, \ref{Eco} and \ref{Chall} study the life cycle, ecosystem and challenges of threat hunting. Section \ref{Fut} develops a future roadmap for future research on threat hunting and lastly, Section \ref{Conc} concludes the paper.

\section{Existing Surveys}\label{ExSurv}

The literature comes with many surveys related to threat management \cite{bbb}. However, some of them or too outdated for such a fast-moving research area \cite{Surv-Conf003}\cite{Surv-Jour001}. Others do not focus on threat hunting \cite{RKDO-Conf001}. Some of existing surveys fail to develop a future road map \cite{AI-Surv-Conf001}\cite{Surv-Jour003}. Most of them do not discuss the role of AI \cite{Surv-Conf004}. Moreover, some relevant surveys discuss threat hunting in a specific area \cite{Surv-Jour002}. These shortcomings motivate our work in this paper.

\subsection{Surveys on Threat Management}

This section explores the topics covered in the existing research of the threat hunting life cycle. The sections explore some recent developments in threat management, hunting, and the role of AI \cite{a4, a5}. Subsection 2.3 includes a table that directly compares the threat hunting features noted in the papers discussed in this section (\textcolor{black}{Table} \ref{tab:threat survey table}).

The article \cite{RKDO-Conf001} stated that with the rapid development of technology, companies are constantly competing to have an advantage over their competitors \cite{RKDO-Conf001}. Because of this change, some companies ignore their duty to consider the security measures that should be taken, to prioritize production speed \cite{RKDO-Conf001}. The researchers enforced that it is crucial for companies to consider and address all security implications that come with their new technologies \cite{RKDO-Conf001}. Threat management concerns have been investigated and solutions have been proposed by industry researchers. 

Multiple studies have initiated the partitioning of resources for threat hunting and mitigating is a somewhat new sector for businesses to consider \cite{Surv-Jour003}. The authors of \cite{AI-Surv-Conf001} developed a deep learning model systems that will allow for constant moderation and quick detection of potential threats in their systems.  Deep learning tools will assist in managing the rapid technological development and the consistently growing attack surfaces that create challenges for modern security frameworks \cite{Surv-Jour003}. When faced with managing the security for a company, there are many approaches to take and perspectives to consider, which will be discussed in this section and visualized in table \ref{tab:threat survey table}.

The management of security in companies has become a paramount focus for most, which has led to the requirement of security reports from companies \cite{Surv-Conf001}. In \cite{Surv-Conf001} it was observed that many of these reports are failing to provide complete and thorough reports on the attack trends because they are strictly based on internal data. These researchers used the data from some of these reports a meta-analysis was completed to identify common trends of malware and ransomware attacks \cite{Surv-Conf001}. Combining these incomplete reports, the researchers were able to provide valuable information that may be used by companies to assist in formulating risk models and estimating potential losses post-attack \cite{Surv-Conf001}.

When considering attackers, understanding the approach of common Advanced Persistent Threats (APTs) is highly beneficial to security \cite{a6}. The authors of \cite{BURM-Conf001} determined that each APT has a defined target and the campaigns are typically launched by an established organization. They emphasized that security professionals need to understand how these intrusion methods are executed and how to detect them \cite{BURM-Conf001}. These sophisticated threats require more than normal IDSs as the attackers are highly knowledgeable \cite{BURM-Conf001}. The authors concluded that organizations need to understand the stages and aspects of APT to prepare for the attack and ensure that they can identify the threat in the early stages \cite{BURM-Conf001}.

The number of “things” connected to the internet is on the rise. More devices are merged through the internet to make day-to-day life easier \cite{RKDO-Conf001}. When considering the Internet-of-Things it is difficult to fathom the extent of high-value data traffic that is transmitted every day. The authors of \cite{RKDO-Conf001} stated that as devices are added to the internet, the threat potential also increases. Protecting this data from threats is a colossal issue that security professionals, corporations, and governments are responsible for keeping \cite{Surv-Jour003}. The research completed in \cite{Surv-Jour003} has examined the many layers in the internet of things (IoT) (perceptual, network, and application) and applied their attributes to a variety of datasets to understand the threats that each level may attract or be vulnerable to. Suggested problems to consider have also been proposed. The authors of \cite{RKDO-Conf001} listed these suggestions as the efficacy of threat detection in the IoT, the implementation of machine learning (ML), and the development of standardization.

\subsection{Surveys on AI-Assisted Threat Management}

The authors of \cite{AI-Surv-Conf001} have recognized the daunting task of leading Insider Threat Detection for organizations. It was understood that many malicious activities can go undetected with the different privileges associated with users, and the rate that digital footprints can get lost in large networks \cite{AI-Surv-Conf001}. These researchers have compiled datasets to provide a clear understanding of Insider Threat Detection in Deep learning \cite{AI-Surv-Conf001}. In \cite{AI-Surv-Conf001} it was discovered that Deep Learning solutions have the ability to enhance the capabilities of Log based Anomaly Detection when processing large sets of data. The CERT Insider Threat Dataset, along with others were used in the research to test a Deep Belief Network (DBN), Autoencoder approach four Recurrent Neural Networks (RNN), and a Convolutional Neural Network (CNN) \cite{AI-Surv-Conf001}. Each of the approaches had unique benefits and downfalls, but ultimately the researchers suggested to transition from academic to industrial solutions \cite{AI-Surv-Conf001}. To do this, the researchers suggested to integrate the Elasticsearch-Logstash-Kibana (ELK) toolset into Deep Learning Models \cite{AI-Surv-Conf001}.

\subsection{Surveys on Emerging Threats}

Cyber intrusions are evolving exponentially as computer systems advance \cite{a7, Surv-Conf003}. The articles \cite{Surv-Conf003}\cite{Surv-Jour001}\cite{SMRQ-Conf001} all observed a consistent rise of threats and vulnerabilities that are constantly challenging the updated computers, mobile phones, and other products. Specifically,\cite{Surv-Jour001} stated that threats faced today go beyond the well-recognized email spam, and have advanced to more complex threats such as botnets, or ransomware. \cite{Surv-Conf003} clearly stated that the current threats have led to a requirement for modernized intrusion detection technologies supported by current research. Attackers have also been observed to still be using older, simple tools for intrusion as well as the modern, advanced threats, so security professionals require a more diverse and tiered knowledge on how to approach the potential threats \cite{Surv-Jour001}. The article \cite{SMRQ-Conf001} highlighted that some threats are so discreet that many times they are not recognized until they have fully infiltrated the target network.

A common emerging threat called Denial of Service (DoS) is a type of threat that exploits a network by overwhelming the contact point so that the network or server is left impaired  \cite{Surv-Conf004}. These attacks are usually used to attack in-demand services such as online banking, streaming, and social media  \cite{Surv-Conf004}. The article \cite{Surv-Conf004} noted that as the internet matures, new layers are being introduced which require modern security methods. The researcher explained that the transport layer provides an opportunity for attacks . They further suggested that the current tactics for security from DoS attacks are still valuable but modernizing them to account for new paradigms will ensure strength  \cite{Surv-Conf004}. For example, the article \cite{Surv-Conf004} encourages using cloud environments as they provide a cost-effective DoS evasion technique.

In \cite{SMRQ-Conf001} it was stated that mobile device security is important to consider as smartphones become more connected and integrated into our lives. The researchers state that mobile security can be improved by ensuring that care is taken to use verified tools when visiting websites and downloading applications. In \cite{SMRQ-Conf001} it was expressed that malicious files can be stored in third party apps and unauthorized websites that attackers have built to look like commonly used platforms, so moderation is extremely important.

Individual phone numbers are linked to each phone and are a very important piece of Personally Identifiable Information (PII) \cite{Surv-Conf002}. Due to the reach of phone numbers, compared to other PII such as email addresses, the article \cite{Surv-Conf002} explains that phone numbers are more enticing for attackers to target. \cite{Surv-Conf002} further discusses that it is somewhat easy for attackers to generate large sets of valid phone numbers that can be exploited later for malicious activities. \cite{Surv-Jour001} says  that every level of threat, simple or advanced, can have catastrophic effects on the target so it is important to be aware of the potential attacks and how to be cautious. As the smartphone landscape increases, security professionals and average users will have to aim to recognize and mitigate attacks \cite{Surv-Jour001}. \cite{Surv-Conf003} noted that any organizations and individuals use Network IDSs to protect themselves against attacks. The users in \cite{Surv-Conf003} write extensive detection rules to maximize the detection system's effectiveness. To best prepare for these attacks,the authors emphasized the importance to understand the attacks by making a comprehensive dataset with an array of metrics for testing and decision making \cite{Surv-Conf003}.  

\subsection{Surveys on Threat Hunting}
Threat hunting is important for all types of technologies. In \cite{Surv-Jour002} it was identified that there is a lack of research regarding non-Windows operating systems. There is constant development of hunting techniques for Windows malware, but there is a lack of work regarding Linux protection \cite{Surv-Jour002}.  The article  \cite{Surv-Jour002} explained that IoT devices are typically run using Unix-based architectures so there is a high risk of exploitation due to researchers’ lack of preparation. Since IoT devices are so integrated into our homes and lives, there could be severe repercussions after an attack \cite{Surv-Jour002}. In \cite{Surv-Jour002} it was stated that ML-based threat hunting has been overwhelmed by the big data problem of IoT devices. The authors proposed a creation of a new platform and CPU instruction set, along with an updated taxonomy of developments could be the solution to fulfilling this gap of protection  \cite{Surv-Jour002}.  The work \cite{Surv-Jour002} also stated that a comprehensive analysis of current research to hone in on the lacking fields for future research is required.

\textcolor{black}{Table \ref{tab:threat survey table} shows the features of existing threat hunting surveys in regards to the topic of threat hunting, and the main discussion points involved.   }

\begin{table}
\centering
\caption{\label{tab:threat survey table} \textcolor{black}{A Summary of Existing Surveys}}
\begin{tabularx}
{0.5\textwidth} { 
  | >{\raggedright\arraybackslash}X 
  | >{\raggedright\arraybackslash}X
  | >{\raggedright\arraybackslash}X 
  | >{\raggedright\arraybackslash}X 
  | >{\raggedright\arraybackslash}X 
  | >{\raggedright\arraybackslash}X | }
 
 \hline
Works & Year & Threat Hunting & Roadmap & AI & General  \\ 
 \hline
 \textcolor{black}{\cite{Surv-Jour002}} &  2021 & Y & Y & Y & Y  \\
 \hline
\textcolor{black}{\cite{Surv-Conf001}} & 2021 & N & Y & N & N \\ 
 \hline
\textcolor{black}{ \cite{BURM-Conf001}} & 2021 & Y & N & N & Y  \\
 \hline

 \textcolor{black}{ \cite{RKDO-Conf001}} & 2020 & N & Y & Y & N \\
 \hline
 \textcolor{black}{\cite{AI-Surv-Conf001}} &  2020 & N & Y & Y & Y  \\ 
 \hline
\textcolor{black}{\cite{Surv-Jour003}} &  2020 & N & Y & N & Y  \\
 \hline

\textcolor{black}{\cite{Surv-Conf004}} &  2018  & N & Y & N & N  \\
 \hline
\textcolor{black}{\cite{SMRQ-Conf001}} &  2016 & N & N & N & N  \\
 \hline
\textcolor{black}{\cite{Surv-Conf002}} & 2016 & N & Y & N & Y  \\
 \hline
\textcolor{black}{\cite{Surv-Jour001}} &  2014 & Y & Y & N & Y  \\
 \hline
\textcolor{black}{\cite{Surv-Conf003}} &  2011 & Y & Y & N & Y  \\
 \hline

\end{tabularx}%

\end{table}

\textcolor{black}{Indeed, in Table \ref{tab:threat survey table}, the first column shows the reference number of the survey analyzed in that row. The next column (\emph{Year}) is ordered in most to least recent year of publication. The column \emph{Threat Hunting} in Table \ref{tab:threat survey table} indicates if the survey directly discusses threat hunting. The discussion of future possibilities and research suggestions is indicated in the following column (\emph{Roadmap}). Finally the column \emph{General} lists if the survey generally discusses threat hunting.   }

Looking at the content of Table \ref{tab:threat survey table} it can be seen that many of the discussed surveys do not explicitly focus on threat hunting, but rather highlights the general application in the field. Majority of the surveys did propose some suggestions for application and future development of the area of focus. Within the studies there was some discussions of the integration of AI, but there was little direct explanations of AIs role. Overall, the surveys did discuss the general properties of threat hunting, and analyzing the collection as a whole gives good insight to the current work and future direction (\textcolor{black}{Table} \ref{tab:threat survey table}). In Table \ref{tab:threat survey table} it can be seen that the article \cite{Surv-Jour002} provides a comprehensive understanding of general and specific threat hunting topics, and the future of threat hunting with the integration of AI.

\section{The Critical Problem of Emerging Threats}\label{Emerg}

In this section we discussed that With the emergence of cyberthreats, you can no longer control or safeguard network borders \cite{a9}, the supposition is in identifying the breach as soon as possible so you are able to minimize its impact. However, there are some ways to identify new threats including new ML method \cite{a8}.

\cite{Resp-Emerg-Jour001} Studies logic solution for emerging Worms and viruses and botnets indicated that hackers are no longer content just to compromise and control individual computers, as they have now turned their attention to virtual networks of zombie systems, which they use to commit wicked actions \cite{Resp-Emerg-Jour001}. In spite of the fact that botnets are comparatively novel and developing threats, the lawful structure supports strong solutions
\cite{Resp-Emerg-Jour001}. In another paper Emerging Cyberthreat Events in Twitter Streams is discussed\cite{TYT-Conf001}. A new ML and text information extraction method is presented in \cite{TYT-Conf001} to identify new and growing cyberthreat incidents on Twitter. In addition, the offered method give the option for the ranking of cyberthreat events in terms of their significance based on extracting the tweet terms that can be considered as named objects or keywords \cite{TYT-Conf001}.

Another article discusses about Automated discovery of Emerging Network Security Threats.The Internet community confronts significant issues in terms of system and network security. The rise of hi-tech crime poses a threat to the expansion of an online business \cite{Resp-Emerg-Conf001}. The security community has grown and used technologies that allow it to stay up-to-date on new threats \cite{Resp-Emerg-Conf001}. In addition to observing unlawful Internet traffic, data analysis is needed to recognize novel and developing threats among the abundance of unlawful, but identified, traffic \cite{Resp-Emerg-Conf001}.

Additionally,\cite{Resp-Emerg-Conf002} explained a consensus-based approach to assessing the effectiveness of modern security products for detecting and containing emerging threats\cite{Resp-Emerg-Conf002}. With the emergence of cloud computing and cyberthreats today, people and companies alike have realized several things\cite{Resp-Emerg-Conf002}. First, there are no longer any network borders under you can control and safeguard. Second, Due to the essence of threats, they are often dispersed both in time and setting that makes recognition very problematic\cite{Resp-Emerg-Conf002}. Third, rather than supposing you can avoid infections, the working supposition is the speed at which can you notice the breach and how do you most lessen its influence \cite{Resp-Emerg-Conf002}.
It was shown that transportation networks in current society must be harmless, protected, and well-organized \cite{Resp-Emerg-Jour002}. Possessors and operators of these networks are growingly using IoT technologies in order to increase their general success \cite{Resp-Emerg-Jour002}. The conducted research suggested that, to the domain-particular security issues, slight emphasis has been committed that emerge when IoT is applied inside the transport area\cite{Resp-Emerg-Jour002}.

In \cite{BVOS-Jour001} explained that Most network management use traditional rule-based IDSs based on identified attack signatures, which do not identify novel attacks. As a result of the inadequate statistical validation of base truth data, which is used to shape regular network behavior, irregularity detection solutions are known to be inclined to to great false positive proportions. \cite{BVOS-Jour001} presented the scheme, execution, and assessment of Citrus, a new ID structure that can identify and organize hostile behavior using graph-based metrics and ML algorithms to solve developing threats \cite{BVOS-Jour001}.

It was shown in \cite{Resp-Emerg-Conf003} that it is becoming growingly difficult for users to locate used IoT devices and comprehend their goals and potentials due to the rise in smooth combination of IoT sensing and activating devices.\cite{Resp-Emerg-Conf003} Providing a mechanism of mapping the IoT and tackling stakeholder needs is one method. IoT maps may, nonetheless, divulge a number of vulnerabilities that will require to be solve. The STRIDE model was used for two case studies to find possible weaknesses and approaches for addressing them in the IoT maps setting.\cite{Resp-Emerg-Conf003}.

Furthermore, \cite{OQSY-Jour001} argued that Block chains and decentralized file storage systems have allowed a broad scope of novel applications and opportunities due to the extensive acceptance of the new generation of decentralized architectures. In article \cite{OQSY-Jour001} blockchain and the broadly used DFS systems was studied and their main challenges and opportunities, especially their immutability and its effect on General Data Protection Regulation (GDPR) compliance is discussed. 

It was shown in \cite{APMQI-Conf001} Organizations with a high reliability rating (HRO) function in dangerous and safety-critical settings where failure prevention takes precedence over conventional performance measurements and cost productivity. Five key HRO features have been recognized by investigation in the military, air traffic control, and similar domains \cite{APMQI-Conf001}. Additionally, In this research \cite{DFGH-Jour001}, the interpretation errors made by Alexa, the speech-recognition engine that controls the Amazon Echo family of devices is examined.  \cite{DFGH-Jour001} describes a novel type of attacks, called skill squatting attacks that exploit shared misinterpretations made by Alexa, and its security implications \cite{DFGH-Jour001}.

The authors of \cite{GHOSI-Conf001} explained that Virtual hosting for lightweight operating systems is possible with container technology. Multi-tier distributed applications are altered greatly by the mentioned technology coming into the view. There are some security issues still associated with allowing several containers to share a single operating system kernel on a multi-tenancy container cloud service due to an imperfect implementation of system resource isolation mechanisms in the Linux kernel\cite{GHOSI-Conf001}.

It was shown in \cite{YUOF-Jour001} that throughout history, cryptocurrencies have been used by cybercriminals for of the privacy and pseudo anonymity they provide. Collecting datasets to train protective systems to identify and examine these attacks by cybercriminals is a substantial challenge for researchers \cite{YUOF-Jour001}. Authors of \cite{YUOF-Jour001} found that there is a substantial amount of research covering the finding and examination of high produce investment programs and pump and dump attacks.

The authors of \cite{State-Conf103} argued that Firewall technology is an essential first step in safeguarding networks of any complexity or scope against attacks as a result of the growing threat of attacks and malefactor activities. Current greatly spread, active, and varied settings make it particularly problematic to design and manage firewall policies, severely limiting their effectiveness. Hence, it is required to automate the firewall configuration if possible \cite{State-Conf103}.

\section{Threat Hunting: Life Cycle}\label{Life}

\begin{figure}
    \centering
    \includegraphics[width=8cm]{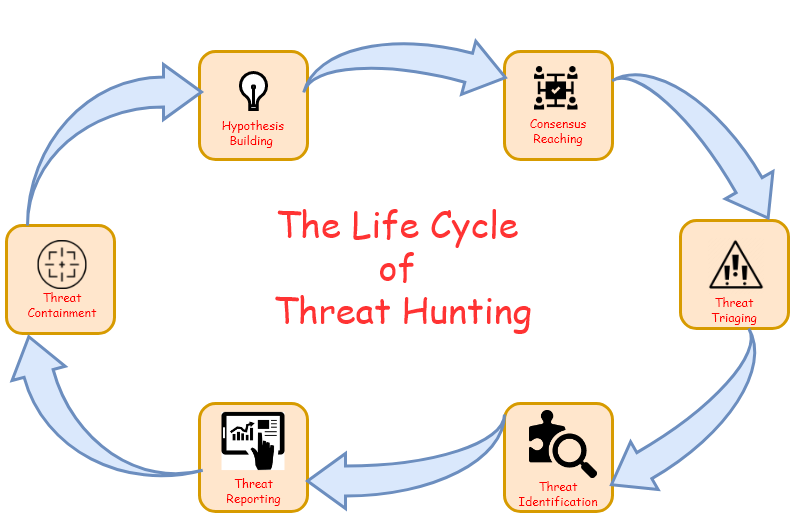}
    \caption{Threat Hunting Life Cycle Stages}
    \label{fig:life}
\end{figure}

\textcolor{black}{Figure \ref{fig:life} shows the five stages of the threat hunting life cycle discussed in this paper, in order of occurrence.}

\textcolor{black}{The threat hunting life cycle is compromised of five stages that can be seen in \ref{fig:life}. The first stage is \emph{Hypothesis Building} which includes details on the development of arguments regarding threat hunting (Section \ref{hypothesis}).  The next is \emph{Consensus Reaching} where cybersecurity professionals analyze and classify the threat (Section \ref{consensus}). The next step, \emph{Threat Triaging} involves planning and delegating tasks to  manage the treat in the best possible manner (Section \ref{triage}).  \emph{Threat Identification} is the next stage and includes threat emulation and threat quantification. More information can be found in section \ref{ID}. \emph{Threat Reporting} includes the processes of threat warning, threat modelling, threat recording, and threat visualization (Section \ref{report}). Finally, \emph{Threat Containment} includes threat tracking and monitoring (Section \ref{contain}). The life cycle starts with hypothesis building and cycles through each stage until the threat has been successfully hunted and defeated, then the process starts again as there are constant emerging threats. }

\subsection{Hypothesis Building}\label{hypothesis}
Threat hunting life cycle is 6 steps including hypothesis building, consensus reaching, threat triaging, threat identification, threat reporting and threat containment. In this section we investigate the importance of each step and examples of using them. \textcolor{black}{
Threat hunting is scientific in nature and begins with steps in the threat hypothesis phase that are designed to create a logical argument regarding an existing threat, next follows with steps in the threat hunt phase intended to validate the argument \cite{Threat-Hunt-Conf012}. Threat hunting hypothesis must create a correlation and causal relationship between a threat and an asset and stick to the scientific method for the exercise to be defined and measured accurately, and produce valuable and repeatable results \cite{Threat-Hunt-Conf012}. By failing to hold to the scientific method, threat hypotheses often present unacceptable or unrelated propositions, which decreases return on investment in cybersecurity defensive efforts because of wasted cycles of threat hunting \cite{Threat-Hunt-Conf012}.}

In \cite{Threat-Hunt-Conf012} paper, Collect Analyze Relate Validate Establish (CARVE) is proposed as a scientific method for developing valid threat hunting hypotheses in the context of a specific organization's information system and environment. In a case study based on the U.S. Computer Emergency Readiness Team (US CERT) technical alert "TA17-293A," the CARVE model is defined as the following: Collect, Analyze, Relate, Validate, and Establish \cite{Threat-Hunt-Conf012}.

 \subsection{Consensus Reaching} \label{consensus}

\textcolor{black}{ 
Consensus identifies new threats as malicious or not when (n/2+1) security products agree on the nature of the threat over time. The method was developed as a simplified extension of the well-known Byzantine Agreement protocol, first discussed by Leslie Lamport \cite{APMQI-Conf002}.}

To test the ability of commercial gateway and endpoint security services to classify and categorize different types of web traffic (malicious content, malicious activity, and non-malicious content), the authors have developed benchmark metrics \cite{APMQI-Conf002}. This methodology was used to evaluate eight gateway protection services for identifying malicious traffic, command and control (C2) communications, and non-malicious content. Consensus is a key component of the methodology\cite{APMQI-Conf002}.

\subsection{Threat Triaging} \label{triage}
\textcolor{black}{ 
Experts can determine whether a suspected malicious file is similar to existing malicious files and triage them accordingly that is one of the quickest ways to identify and assess numerous malicious samples  \cite{Threat-Hunt-Conf004} . Using the most appropriate triaging method can significantly improve the precision of further static and dynamic analyses, as well as saving a great deal of time and effort  \cite{Threat-Hunt-Conf004} . There are currently three popular and proven triaging methods: fuzzy hashing, import hashing, and YARA rules, which you can use to determine whether, or to what degree two malware samples are similar. The mechanisms among these three methods differ significantly, and comparing them is difficult \cite{Threat-Hunt-Conf004}. }

The authors of \cite{Threat-Hunt-Conf004} evaluate three different approaches for triaging four of the most relevant ransomware types: WannaCry, Locky, Cerber, and CryptoWall. The study evaluates their triaging performance and run-time system performance, emphasizing their limitations.

\subsection{Threat Identification}\label{ID}
\textcolor{black}{ 
 Most companies use signature-based commercial antivirus products that do not provide organizations with the sensitivity needed. In addition to antivirus products, ML techniques can also perform a significant role in malware detection\cite{CMLF-Jour001}.} Performance-based malware target recognition is extended in \cite{CMLF-Jour001}, which currently relies exclusively on static heuristic features. In experiments, the architectural component achieved a detection accuracy of 98.5

 Insider threats are more problematic to discover since insiders may know more about an organization's information security policies and procedures than outsiders \cite{CMLF-Conf001-1}. The insiders in an organization have access to their organization's information systems, as well as legitimate functions to carry out that require the use of these systems \cite{CMLF-Conf001-1}.
 In \cite{CMLF-Conf001-1} an insider threat detection prototype is evaluated against a set of experiments that evaluate its ability to detect scenarios that have not previously been considered or seen by its system developers. Without prior knowledge of what scenario is present or when it occurs, this paper shows the capability of detecting a variety of insider threat scenario instances embedded in real data \cite{CMLF-Conf001-1}.

In \cite{Lurk-Conf001}, it is explained that a computer system's integrity depends on the integrity of its kernel code and data. Kernel modifications are tougher to detect than their user-level equivalents. However, the tampering pattern has so far been limited to hiding malicious objects in user space. In this case, kernel data structures are manipulated in order to intercept user requests and change the user’s view of the system \cite{Lurk-Conf001}. Thus, defense techniques are based on detecting such hidden behavior. These kind of attacks are hidden so they damage the system without being understood by the user or IDS, and they states a systemic problem inside the kernel. The recent generation of kernel integrity monitors can't discover this kind of attack without prior knowledge of its signature\cite{Lurk-Conf001}.

The authors of \cite{Threat-Hunt-Conf006} explained that a continuous monitoring of systems is necessary for the threat hunting process to keep indicator of compromise (IoC) and thresholds of normal behavior up-to-date and match changes in the monitored systems. Google's Rapid Response (GRR) is capable of collecting huge numbers of artifacts from a large number of clients in a timely and distributed manner \cite{Threat-Hunt-Conf006}. Nevertheless, the options for exporting the data collected are still under development, and this might pose a challenge for large-scale threat hunting\cite{Threat-Hunt-Conf006}.

Study \cite{State-Conf018} explained that an evaluation index system for radar threat identification is constructed, and a radar threat identification method based on Entropy-TOPSIS is proposed as a method of identifying radar threat quickly and efficiently. The method emphasized the uncertainty of the target attribute and avoided the subjective assumption of the traditional TOPSIS method. By solving the problem of threat sequencing, the method is proven to be valid and feasible \cite{State-Conf018}.

It was shown in \cite{State-Conf019} that navy Research Lab developed MISTI to detect potential threats with gamma rays from a distance. The MISTI system has been used to demonstrate a new technique for localizing sources at standoff using proximity techniques\cite{State-Conf019}.

In \cite{State-Conf020} paper, the authors explained that it is necessary to create security of the system in the early Phases of the software Cycle of Software Development since each year, large amount of money are lost because of bugs in software security caused by inadequate or incorrect security processes. Numerous breaches of security happen on software systems. Many solutions have been suggested in the scientific literature to solve security issues \cite{State-Conf020}.

According to \cite{State-Conf021}, the Traditional security methods are useless toward insider threats. The identification of insider attacks plays a crucial role in detecting insider threats. An effective method to detect an entity that pretending to be another entity is to monitor the user's unusual behavior \cite{State-Conf021}. To build a database that stores user's behavior trait information, this method uses a weight-changeable feedback tree augmented Bayesian network \cite{State-Conf021}. However, the amount of information is massive, and a process information model of user's behavior attribute needs to be established based on dimensionality reduction using rough sets of data \cite{State-Conf021}. When user behavior departs from the characteristic model, the minimum risk Bayes decision can effectively identify the real identity of the user \cite{State-Conf021}.

It was shown in \cite{State-Conf023} that the number of reported security threats hitting organizations has increased recently. It is believed that some of them result from the assignment of inappropriate authorizations on organizational sensitive information to users \cite{State-Conf023}. Therefore, organizations must identify risks as early as possible by identifying the risks resulting from inadequate access rights management, and finding solutions that would prevent such risks \cite{State-Conf023}. The article \cite{State-Conf024} explained that Organizations have reported grow in security threats lately. In some cases, they result from users having incorrect permissions on sensitive organizational data. Therefore, it is essential that organizations identify as early as possible the risks that arise from improper access right management and find solutions to avoid them from occurring \cite{State-Conf024}. The article \cite{State-Conf025} described that collaboration across domains, devices, and service composition is becoming increasingly common in business applications. In addition to its isolated systems, security should focus on the general application scheme, comprising interaction between its units, devices, and services. Security Threat Identification and testing is a toolkit that \cite{State-Conf025} introduce to assist development teams with security testing of their underdevelopment applications with the aim of identifying delicate security logical errors that may go hidden by using existing industrial technology
\cite{State-Conf025}.

It was shown in \cite{State-Jour010} that to capture the distributed digital footprints of malicious insiders among a variety of audit data sources over a prolonged period of time, current methods usually use scoring mechanism to arrange alerts produced from several sub-detectors. Mentioned roaches lead to great deployment complexity and extra cost\cite{State-Jour010}. The authors of \cite{State-Conf026} explained that US Navy with the help of Stottler Henke extends and improves enhance the Intelligent Surface Threat Identification System (ISTIS). ISTIS enhances Littoral Combat Ship (LCS) Surface Mission Module including threat ID process, quality and productivity \cite{State-Conf026}.

As software's importance in modern society grows, so does the threat to it. IT's greatest problem is building software that is invulnerable to these threats\cite{State-Conf027}. By analyzing UML models, this work of \cite{State-Conf027} possibly fills a void of threat identification methodologies, and a void of automated methods for detecting threats based on UML models \cite{State-Conf027}.

Studies in \cite{State-Conf028} noted that radiological threats in city and country are located within +/- 10m in range by Mobile Imaging and Spectroscopic Threat Identification system. The data acquisition system for MISTI was developed using the most up-to-date commercially available hardware \cite{State-Conf028}.

An analysis of the usage of process modeling and insider problem is presented in this work \cite{State-Conf029}. Initially, it is explained the work of a process with process modeling. Next, the agents who are performing specific tasks conduct various analyses to decide the way process may be compromised
\cite{State-Conf029}. Based on \cite{State-Conf030} real-time security risk assessment practices usually rely on IDS alerts as the only source of risk information. As network security becomes more complicated, their assessment outcomes are more susceptible to of false positives. By making use of many risk factors, this paper \cite{State-Conf030} offers an online fusion model for dynamic network risk assessment.

In \cite{State-Conf031} it is mentioned that MISTI, the Mobile Imaging and Spectroscopic Threat Identification system developed to operate in urban and rural settings, is now going through characterization activities\cite{State-Conf031}. MISTI is a mobile origin discovery and imaging system that can locate a radiological source within +/- 10 meters of the location of the source. Data acquisition system developed by MISTI uses the newest commercially beneficial hardware to meet MISTI's needs \cite{State-Conf031}.

The objective of this brief report \cite{State-Conf033} is to present valuable statistics for networked defense analysts including in threat identification on a controlled experiment. In these statistics, the correctness of top-central actor results taken from relational data typically detected in practical data-sets is estimated.\cite{State-Conf033} During the experiment, cellular social networks are included with four types of data errors including missing links, missing actors, extra links, and extra actors
\cite{State-Conf033}.

The article \cite{State-Jour011} explained that In order to neutralize terrorism and espionage efforts, quicker, more precise, and simpler to execute threat identification systems for hidden electronics are required. A new, non-intrusive, repeatable, consistent, expandable, and simple-to-implement recognition and identification system is presented in this paper \cite{State-Jour011} for identifying threats using unintentional radiated emissions (URE).

in \cite{State-Conf034}, Many groups have expanded millimeter-wave and terahertz (THz) imaging systems for hidden weapon detection over the past several years. System design at millimeter-wave span usually provides decent transmission power via clothing materials, however have very limited spatial resolving power at spaces greater than a few meters for practical aperture sizes
\cite{State-Conf034}. The article \cite{State-Conf035} explained that Protecting critical data from being demolished and taken illegally is best achieved by understanding the situation of network attacks, and then detecting the threats. This article \cite{State-Conf035} focused primarily on the new energy plant, and plan to examine network attack scenarios in order to identify all the possible scenarios that may arise. In \cite{State-Conf036}, the Internet-of-Things model have created a vast security gap, as well as opportunities. There are many studies exploring security via device identification, cryptography, and network security protocols, but the problem of whether we can rely on the metrics and data being sent by IOTs, remains a challenge in spread wireless scenarios \cite{State-Conf036}.

Due to the fact that contemporary smart grid operation is greatly depend on spread microprocessor based control, there is a necessity for interoperability standards to deal with varied data in smart grids. \cite{State-Conf037} starts the article by studying the Sampled Measured Values Protocol and its advantages, and it then analyzes its vulnerabilities and identifies the linked cyberthreats. Next current security measures is outlined and, lastly, whether neural network predictor can identify spoofed samples is explored \cite{State-Conf037}.

It was shown in \cite{State-Conf038} that administrators face numerous challenges in securing heterogeneous and complex networks that can address by the tool called Security Information and Event Management to control and detect the threats. \cite{State-Conf038} solve the issue of performance by offering a Latent Semantic Analysis in order to lessen the redundant noise in an enormous data produced from devices\cite{State-Conf038}. The authors of \cite{State-Conf039} explained that Cyberthreat intelligence efforts are centered on internal threat feeds such as antivirus and log files. Although this approach is beneficial, it is reactive and depend on practice that has already taken place.\cite{State-Conf039} Organization can better safeguard their infrastructure and offer enhanced CT, if they learn about malicious hackers prior to an attack\cite{State-Conf039}.

In \cite{State-Conf040},The Cyber-Physical System contains a mixed combination of physical and computer components that are usually watched by computer-based algorithms. \cite{State-Conf040} Nevertheless, it has been necessary to protect insiders from penetrating the probable code of conduct in keeping very important data and assets of organizations. Using human brainwaves with applying deep learning algorithm has been displayed to be beneficial in \cite{State-Conf040} identifying threats to create attacks in critical infrastructure. Paper \cite{State-Conf041} explains that businesses face a significant threat from malicious emails. To protect against email threats, such as targeted attacks, traditional signature, rule-based email filters and advanced sand-boxing tools each have their own shortcomings. The article of \cite{State-Conf041} offers a predictive analysis method that achieves detection and forecast on unseen emails productively uses static analysis and ML to distinguish legitimate from malicious emails.

According to \cite{State-Conf042}, in order to specify the rate of insurance coverage more accurately, the organization's assets must be specified. These parameters probably demonstrated by indicators to plan the influence of particular cyberthreats on an organization's information systems. It is essential to model the communication between parameters and cyberthreats based on parameters that are important at the start of algorithm \cite{State-Conf042}. In paper \cite{State-Conf043} it was noted that The National Oceanic and Atmospheric Administration has been managed to start proactively assessing catastrophic oil and other chemical releases from soaked sources following new events in our National Marine Sanctuaries and throughout our country. Data from federal, national, and private sources are gathered in the Resources and UnderSea Threats (RUST) database, which is used to list this possible threat and specify its range via analysis\cite{State-Conf043}. Based on a logical probabilistic method on a collection of security properties which consider the details of botnet attacks, a method to identify and act against the negative impacts of a botnet using estimates of the risks of botnet attacks exist for any object-risk business network
\cite{State-Conf044}. Study \cite{State-Conf045} demonstrates how three closely joined swarming pattern analysis designs including profiling, clustering, and forecasting improve each other's results greatly. It also indicates that systematic assessment experiments approve the research hypothesis
\cite{State-Conf045}.

According to \cite{State-Conf046}, Mobile devices become targets of financial gain attack because they keep confidential personal information including credit cards and passwords. In \cite{State-Conf046} study, they identify threat patterns in order to detect mobile malware. Using the suggested method, malicious behavior on Android mobile devices can be identified by analyzing function calls and data flow. It was shown in \cite{State-Conf047} that politicians and healthcare providers are concerned about electronic protected health information. Healthcare providers have to ensure information security in because of the growth in data breaches and the cost linked with them. The study \cite{State-Conf047} found that transitive information risks have significant consequences for healthcare organizations and supervisors. Information security in the healthcare setting will be considerably boosted by detecting these risks\cite{State-Conf047}.

The study \cite{State-Conf048} was conducted to recognize visible events associated with insider sabotage. Almost 71\% of the cases they studied did not have a noticeable malicious action before attack.Earlier to the attack, most of the events detected appeared to be behavioral, not technical\cite{State-Conf048}. The installation of software onto the target company’s IT systems accounted for approximately 33 percent of the detected technical incidents before an attack
\cite{State-Conf048}. Using an outcome-based learning model to identify emerging threats, the authors in \cite{State-Conf049} introduce simulation and experimental results. In order to provide a framework for the study of emerging threats, this model contains judgment, decision making, and learning theories
\cite{State-Conf049}.

The research in \cite{State-Conf050} indicated that, Often times, managers and coworkers observed signs of stress, dissatisfaction, or further issues on the part of insider criminals but did not raise an alarm. Psycho-social sings are difficult to use because the indicators are not identified and behaviors are not recorded as a result they cannot be evaluated\cite{State-Conf050}. To evaluate employee behavior connected with higher risk of insider malbehavior, a psycho-social model was established \cite{State-Conf050}. In \cite{State-Jour012}, the widespread use of wireless networks in daily life makes them an important attack target for criminals. In \cite{State-Jour012} paper, a methodology is presented for the systematic identification of vulnerabilities connected with wireless access protocols, as well as a quantitative assessment of the consequential risks for mobile operators by using attack trees taking into account existing legal structures.\cite{State-Jour012}. A biometric of intent (BoI) is based on AI that offers a novel method to biometric identification. Based on the analysis of facial expressions, \cite{State-Conf051} proposed BoI framework that allows law enforcement agencies to use a systematic preventive security method designed to lessen the likelihood of illegal attacks by malefactor individuals by understanding their emotional state\cite{State-Conf051}.

Study \cite{State-Conf052} stated that the transitivity threat is when an unrelated action reveals information to an unintentional audience. A transitivity threat related to social networking sites is when automated transmission of data occurs. As part of \cite{State-Conf052} study, they model social network content, friends, friendship relations, and privacy policies as access permissions to content. In \cite{State-Conf053}, Various government agencies and organizations are only start to take advantage of the great potential of visualization for stopping, identifying, and reduce security threats. The article \cite{State-Conf053} used classifications and visualization approaches of insider behavior to create a pattern of satisfactory actions based on workgroup orderings \cite{State-Conf053}. The authors of \cite{State-Conf054} stated that religious extremism supports violence in the service of God, including killing. The paper \cite{State-Conf054} used a method to forecast future threats involving religious extremism in Sri Lanka. In this study, a ML model and opinion dictionary were taught using cautiously selected social media text data, and each text was categorized into religious-extreme
\cite{State-Conf054}. The article \cite{State-Conf055} explained that We are arriving a time when critical services and applications will be reliant on on 'coalitions of systems' due to the development of cloud computing and system-of-systems. CoS are a form of system resembling to systems-of-systems, except that they focus on covering self-benefits instead of a broad mission\cite{State-Conf055}.

It was noted in \cite{State-Conf056} that a cloud computing infrastructure comprises multiple virtual machines running on a host, which is a physical platform. The virtual machines are checked and controlled by software based on kernels such as hypervisor or Virtual Machine Monitor (VMM) \cite{State-Conf056}. The vulnerabilities in VMMs make them susceptible to attacks that may be carried out by insiders or outsiders. The virtual trusted platform module, trusted virtual domains, and virtual firewalls are all security methods that must be employed to safeguard a secure virtualized cloud computing infrastructure \cite{State-Conf056}. In \cite{State-Conf057}, sensitive information are shared through online discussion forums and other platforms. Attackers can exploit this information in order to attack critical infrastructures. Many of the studies on the hacking of computer networks have emphasized on improving classification of cyberattacks but have ignored the exchanging information between the related actors. The paper \cite{State-Conf057} used automated analysis tools to examine the language of the attackers and detect possible threats for critical infrastructures.

Runtime verification methods analyzed in the paper \cite{State-Conf058} as a means of identifying and solving to cyberthreats. For this purpose, it examines the effectiveness of runtime verification for novel threats and discusses the specific use of mentioned methods by state actors. \cite{State-Conf058}. In article \cite{State-Conf059}, it is showed that Attacks by insiders have the capability to lead to severe consequences, financially, reputationaly, and even complete breakdown of the company. The paper \cite{State-Conf059} present a method that comprises several views, including a tool that detects irregular activity of users and a plan that makes user and role behavior visible over time. The article \cite{State-Conf061} explained that Communication and collaboration tools that are facilitated by technology offer numerous advantages, but there are also some drawbacks. Online malicious behaviors can undermine the usefulness of these tools. A research \cite{State-Conf061} of the antecedents of online deviant behavior indicated the likelihood of computer users being hostile to others increases when they act without fore though and feel guilty \cite{State-Conf061}.

\subsubsection{Threat Emulation}

\textcolor{black}{
The use of this approach help organizations to discover advanced attack mechanisms and measure their ability to detect attacks \cite{an1}. As opposed to traditional approaches that emphasize on identified threats including vulnerability assessment and penetration testing, new unknown threats can be identified and addressed with this method\cite{State-Jour017}.}

In addition, the authors of \cite{State-Jour020} stated that When dealing with APT, defenders must detect the area where an adversary is spread as soon as possible. The discovery occurs as part of an incident response operation called Threat Hunting, during which defenders identify attackers within the compromised network\cite{State-Jour020}.

\subsubsection{Threat Quantification}

Regarding threat quantification, we can mention to \cite{State-Conf087}, this model can be used to describe complicated network attacks \cite{State-Conf087}. It defines the threat of an attack and the quantization method of each index in order to introduce complexity and harmfulness of network attack. Next, it proposes a method to analyze network threats that is not target-oriented\cite{State-Conf087}. Additonally, In \cite{State-Conf088} explained a great number of novel and different attacks happen frequently, and inadequate security experts and tools make it problematic to examine and address them. \cite{State-Conf088} provided an approach of analyzing the threat of IoC for cyber incidents, and of calculating its value as a measureable value to check the precedence of cyber incidents that happen in large numbers \cite{State-Conf088}.

\subsection{Threat Reporting} \label{report}
Threat reporting includes four processes; threat warning, threat modelling, threat recording, and threat visualization. In the following we address them.

\subsubsection{Threat Warning}
One of the application of this threat warning is in hit avoidance system. In this regard, \cite{State-Conf089} introduces a signal processing chain for a double hand, infrared, imaging threat warning system with small degree processing for clutter suppression. There are steps in the system that carry out frame registration, adaptive clutter suppression, adaptive threat discovery, taxonomy, and tracing \cite{State-Conf089}.

\subsubsection{Threat Modeling}
\textcolor{black}{
Threat modelling has fundamental three practices. First is identifying information and services that are necessary for the system. Second is creating a summary about how assets are kept and processed \cite{State-Conf062}. And finally, identifying threats that impact the identified system assets. These important activities make software engineering secure \cite{State-Conf062}. }
The purpose of \cite{State-Conf106} paper is to provide a theoretical framework to model frameworks or sample-based attacks that are broader than a single exploit deployed against a solo target. Methods comprising Cyber Kill Chain, STRIDE and the MITRE ATT\&CK structure model actions can be used to attack or stopped to secure organization have various levels of detail \cite{State-Conf106}.

\subsubsection{Threat Recording}
Threat recording happens frequently when cybersecurity practitioners are confused with tackling cyberattacks because there is not adequate attack-defense mapped framework to safeguard their systems and network from threats \cite{an2, State-Conf104}. Cyberthreat dictionary provides immediate practical solutions and methods to address this issue by mapping MITRE ATT\&CK Matrix to the NIST Cybersecurity framework
\cite{State-Conf104}.

It was shown in \cite{State-Conf105} that the information which is provided by the MITRE ATT\&CK including attackers’ tactics, techniques, and procedures would be very useful to diagnose and mitigate attacks. Using ML analysis on APTs and generated reports by this framework regarding software attack is able to predict new attack techniques based on the old ones
\cite{State-Conf105}.

\subsubsection{ThreatNew technology cause many applications to arises Threat Visualization}
\textcolor{black}{
Visualization is an important tool that help security analyst to safeguard modern organizations \cite{State-Conf091}. THACO is an open source threat analyst console that adapt to DNS-based network threat analyst requirements through using scalable visualization technique, a multi-grouping, zoomable treemap\cite{State-Conf091}.}
In \cite{State-Conf090} It is stated that the newest item into the congested field of imaging technologies is THz imaging. The T-ray has greater capacity in the area of concealed objects detection than X-rays because it is not dangerous to humans. Due to the poor quality of THz imaging systems, it is necessary to integrate them with the high-resolution images from a vision camera. By using THz and VIS cameras the authors of \cite{State-Conf090} aimed to create a system safe to humans that can identify concealed objects. Based on THz and VIS cameras, \cite{State-Conf090} introduced a multispectral passive imaging system for picturing of concealed threats.

According to \cite{State-Conf092} as a result of current developments in computing, communications, software, and hardware technologies, the IoT has grown beyond its state of beginning and is a subsequent advance technology in changing the Internet into a completely unified Future Internet.\cite{State-Conf092} introduced a new threat visualization tool for wireless sensor networks which is called VisIoT. It is a visualization tool with human interactive capability that can monitor and find anomaly in systems. It also can detect destroying security attacks such as wormhole attacks and Sybil 
\cite{State-Conf092}.

The article \cite{State-Conf093} stated that decreasing the processing time of data is one of the most demanding tasks in the arena of information security. Precise data visualization enhances the analysis process by decreasing the data processing time that is the most demanding task in the information security. The visualization technique introduced in \cite{State-Conf093} contains two and three dimensional demonstration of the threat model.

In \cite{State-Conf094}, The As cyberattacks are becoming more complicated, identifying and mitigating them in a timely manner is becoming increasingly difficult. An innovative cyberthreat detection and visualization platform was offered in \cite{State-Conf094} .A version of it is accessible and being used in real-life scenarios: it is a cyberthreat platform that gathers 107 million malware events from different data sources and deliver visualization and alerts in real-time for more than 2.7 million of infected unique IPs distributed all over the world \cite{State-Conf094}.

\cite{State-Conf095}
Toward a visualization-supported workflow for cyber alert management using threat models and human-centered design  

In  \cite{State-Conf095}, tools are required for cyber analysts to assist them combine the data they previously have and support them to create proper minimum point in opposition to which for comparing anomalies. In addition, existing threat models, which cyber analysts often use to form their research, are rarely incorporated into support tools. The authors of \cite{State-Conf095} describes their work with cyber analysts to comprehend logical process and how one model, the MITRE ATT\&CK Matrix, is used to form their logical thinking. As part of their threat model design, they attempt to map particular data required by analysts into their visualizations  \cite{State-Conf095}.

The integration and distribution of modern networking systems, applications, and services \cite{an4} has become more complicated, making them more difficult to manage and safeguard \cite{State-Conf097}. In this regard, \cite{State-Conf097} proposed an approach that uses attack graphs and layered security method to create attack scenarios. It focuses on threat identification and helps with the decision making practice.
\cite{State-Conf097}.

In \cite{State-Conf098}, daily network and security operations require the ability to identify typical, malicious, abnormal, and unanticipated behaviors in routing update streams. \cite{State-Conf098} explained Bigfoot, a Border Gateway Protocol (BGP) update visualization tool, which is developed to emphasize and evaluate a kinds of behaviors within update streams. IP geo-location is fundamental to Bigfoot, which visualizes the announcement of network prefixes. Various representations of polygons for network footprints are examined in \cite{State-Conf098} and how simple implementations of IP geo-location can lead to representations that are hard to understand is demonstrated\cite{State-Conf098}.

\subsection{Threat Containment} \label{contain}

Threat containment includes threat tracking and monitoring. When malware breaches an enterprise, incident responders need to be on the lookout and act fast to contain the threat until it can be eradicated from the environment.

According to \cite{State-Conf099} the development of robotics has considerably augmented the difficulty and amount of issues that groups of robots can resolve. The purpose of \cite{State-Conf099} is to resolve a multi-threat control issue by using alike and autonomous robots that create dynamic teams. An analysis of methods that use and do not use wireless communication is presented by an emphasizing on the impacts of using wireless \cite{State-Conf099}.

Based on \cite{State-Conf100} It is essential to have an elastic and scalable factory network in order to deal with the growing number of devices and causing traffic. This can be understood through softwarization technologies including Network Function Virtualization (NFV).\cite{State-Conf100}  developed their prior work to demonstrate threat detection by building an NFV-based on-premises IDS that is combined into their industrial-specific network services.\cite{State-Conf100}.

Threat-aware deployment of sensors and systems of robots can work in pair to identify, evaluate, quarantine and hold threats. \cite{State-Conf101} presents a model, a scheme and a categorized architectural operation of a threat detection system. Using a convergent architecture,\cite{State-Conf101} propose a varied set of operationally independent systems, ranging from in situ sensors, sensor robots (mobile sensors) to aerial reconnaissance sensors, each of which is capable of working in combination. As a way to improve the handling of a threat,\cite{State-Conf101} suggest a deployment strategy arranging in order of rank, which is especially attentive to data integrity and false alarm lessening \cite{State-Conf101}.

An adversarial formulation is used in \cite{State-Jour027} to inspect cyberthreat spread over networks. The authors of in \cite{State-Jour027} suggest an analytical framework to explain the accidental dynamics of cyberthreat spread in varied sub-networks, based on Kendall's birth-death-immigration model. The problem has two formalizations, which are both based on zero-sum games between two adversaries: an adversary proceeding cyberthreats through the different sub-networks, and a defender delivering countermeasures to lessen the threats
\cite{State-Jour027}.

The problem of modeling and having several cyberthreats spreading through many subnets of data network is inspected by \cite{State-Conf102}. This work made use of the Birth-Death-Immigration model to present that the features of this model can be exploited to offer best resource allocation through the attacked subnets
\cite{State-Conf102}.

\subsubsection{Threat Tracking and Monitoring}

The threat monitoring process or solution is responsible for constantly monitoring among networks and/or endpoints for signs of security threats such as efforts at intrusion and data exfiltration \cite{a9,a8}. Monitoring threats gives IT professionals the ability to monitor the network and the users who access it, allowing them to take stronger measures to safeguard data and stop or reduce the damages caused by breaches.

In \cite{Threat-Hunt-Conf005} ransomeware is one of the best attack vectors for threat actors that has been used for financial benefit. It has resembling patterns in their malicious code that can be helpful for identifying them. To identify the source of the attack, first, features and signatures of great quantity of malware samples should be gathered. \cite{Threat-Hunt-Conf005} paper offers a productive fuzzy analysis method to collect ransomware samples.

Cyberthreats are found by an Internet cyberthreat monitoring system by making use of network sensors used on the Internet at a specific points\cite{State-Conf096}. This system examines factors regarding attacks including time, source, type and next creates a visual analysis of the result\cite{State-Conf096}. Currently, existing systems only display statistics by country or hourly fluctuations in attacks. The use of these systems makes it problematic to recognize the adversary source, spreading, and relationship between the origin of the attack and the target
\cite{State-Conf096}.

\subsubsection{Vulnerability Hunting}
The potential financial losses incurred by smart contract vulnerabilities have raised a lot of concern\cite{Vuln-Jour001}. It has been confirmed that matching-based finding methods extrapolating recognized vulnerabilities to unknowns can be effective on other platforms. A direct adoption of the technology to smart contracts is, however, hampered by two issues, namely, variety in byte-code generation causing from fast development of compilers and interference of noise code simply produced by the uniform business logic. As a solution, the author of \cite{Vuln-Jour001} propose byte-code-oriented standardization and part techniques to enhance byte code matching
\cite{Vuln-Jour001}.

\section{Threat Hunting: Ecosystem}\label{Eco}

As technology expands into more commonly used tools the ecosystem for potential attacks also grows. Each of these new integrations has associated vulnerabilities. Because of these threats professionals are working to develop the correlating security tools. 

Subsection 5.1 discusses the real world tools that are being converted to 'smart'. This subsection also discusses some of the newly recognized associated risks. Subsection 5.2 then explains the areas of defense that can be used to protect these new devices, focusing on specific techniques and tools. The final subsection 5.3 explains the importance of hunting for the attack and some associated methods.

\begin{figure}
    \centering
    \includegraphics[width=8cm]{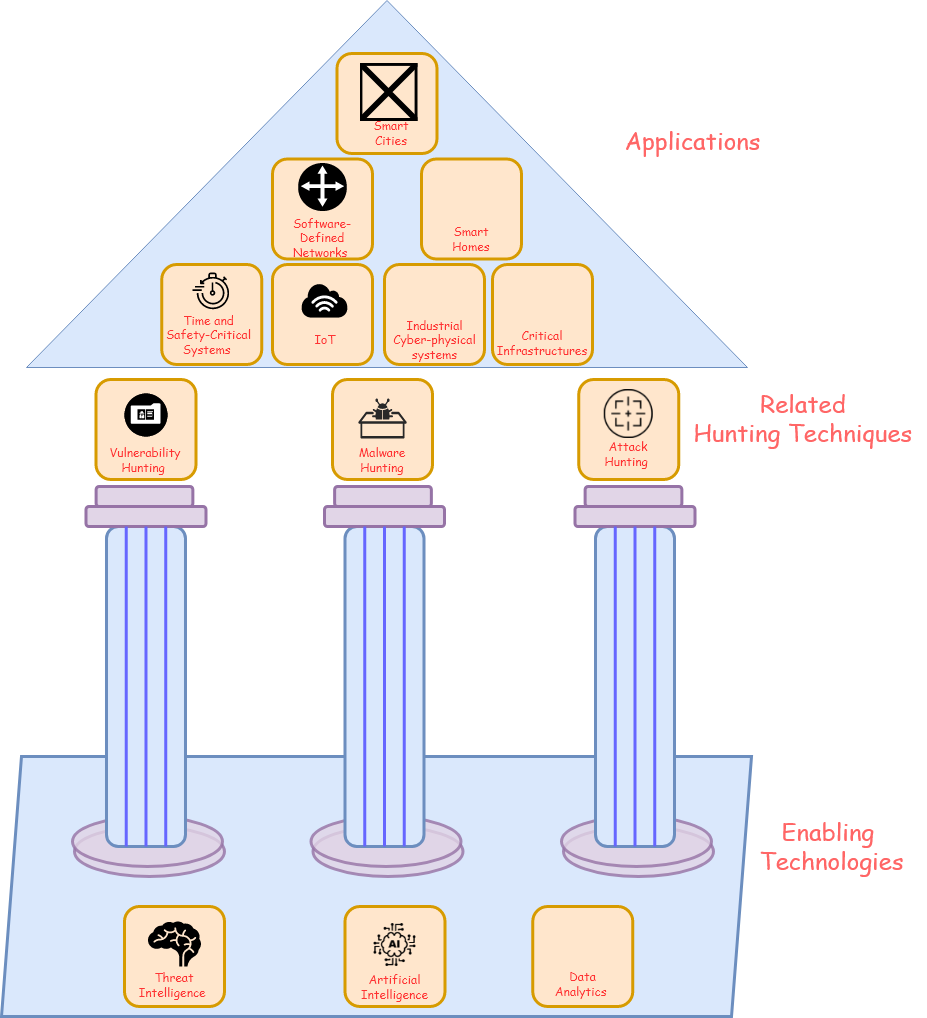}
    \caption{Ecosystem of the Current Threat Hunting Landscape}
    \label{fig:eco}
\end{figure}

The threat hunting ecosystem is compromised of the \emph{Enabling Technologies} and the \emph{Application} of those technologies. The ecosystem is also complimented by the \emph{Related Hunting Techniques} which can all be seen in Figure \ref{fig:eco}. 
The enabling technologies include AI, data analytics, and threat intelligence and are discussed in section \ref{tech}. The application of threat hunting can be seen in smart homes, smart cities, IoT, industrial cyber-physical systems, Windows and Android systems, time and safety-critical systems, software defined networks, and critical infrastructures (Section \ref{application}). These technologies and applications are accompanied by hunting techniques including malware hunting, attack hunting, and vulnerability hunting, which are discussed in section \ref{related}.

\subsection{Applications}\label{application}

\subsubsection{Smart Homes}

With the development and integration of technology into everyday life there has been more integration of computers into homes \cite{an3}. These new ‘smart home’ features make users life easier and enhances the functionalities of the home \cite{Threat-Hunt-Conf016}.\cite{Threat-Hunt-Conf016} identified that IoT devices collect a huge amount of data that is constantly being transported between the gadgets and the homeowner, which requires analysts to formulate solutions to overcome the current security limitations \cite{Threat-Hunt-Conf016}. They explained that threat hunting can be used to understand the vulnerabilities of these gadgets, which will allow researchers to gain a better understanding for responding to new threats and mitigating breaches \cite{Threat-Hunt-Conf016}. The article \cite{Threat-Hunt-Conf016} identified that the best way for analysts to hunt for threats is to initiate a cognitive middle ware for cooperation with the homeowners’ gateways. This connection would allow for automatic reconfigurations to prevent threats as well as mitigation and remediation in the event of an attack \cite{Threat-Hunt-Conf016}. The article  \cite{Threat-Hunt-Conf016} concludes that the assurance of privacy is ensured for homeowners’ as well using this method with the use of two stage concealment protocols \cite{Threat-Hunt-Conf016}.

\subsubsection{Smart Cities}

Smart cities have been proposed in new development plans. Software-Defined Networking (SDN) \cite{an5} has changed the landscape of networking and has encouraged the success of efficiently dealing with network resources and initiating programmability \cite{Master-Thes003}\cite{an6}. \cite{Master-Thes003} explained that flexibility and adaptability are achieved by SDN by compartmentalizing the control and data of the environment.  \cite{Master-Thes003} explained that when SDN infrastructure is combined with ML models an elevated threat hunting system is created. This system allows for automatic handing of threats such as lateral movement with a high level of accuracy \cite{Master-Thes003}. With the development of large ‘smart cities’ there will be an exponentially high demand for security solutions \cite{an7}. Using an intelligent threat hunting system will be a crucial asset to managing the large amount of data traffic that will be produced \cite{Master-Thes003}. The article \cite{Master-Thes003} concluded that intelligent system proposed will allow for better network security as constant updates and threat hunting can be facilitated \cite{Master-Thes003}.

\subsubsection{IoT}

The IoT is a diverse ecosystem \cite{an8}. The article \cite{State-Jour019-1} noted that all of the devices that are a part of the IoT environment are targets for attack because of their diverse application. To achieve security of these devices the authors of \cite{State-Jour019-1} proposed a multikernel SVM, which utilizes the gray wolves optimization techniques. This model secures the IoT cloud-edge by optimizing the selection process of determining if applications are malicious or benign \cite{State-Jour019-1}. The study explains that far less training is required compared to the models predecessor is required to achieve meaningful results \cite{State-Jour019-1}. \cite{State-Jour019-1}concluded that the multikernel SVM produces more accurate results with less computational cost and training time.

\subsubsection{Industrial Cyber-physical Systems}

As study was conducted in \cite{Surv-Jour002} to address the problem of industrial cyber-physical systems (ICPs) facing an emerging threats in cybersecurity. An additional research project  \cite{State-Jour016} identified that these threats are due to the large scale and complexity of the systems. The authors of \cite{State-Jour016} developed a federated deep learning model for threat hunting against ICPs that emulates the temporal and spatial frameworks of the network data. The model has been designed to deploy on suitable edge servers and can maintain reasonable resource delegation \cite{State-Jour016}. From  \cite{State-Jour016} the authors determined that privacy of users can be maintained while efficiency of the program is improved.

\subsubsection{Windows and Android Systems}

As previously discussed in section 5.1.3 IoT devices are facing many attacks \cite{State-Jour019-1}. The researchers in \cite{State-Jour019-1} created an SVM to optimize model training. This method can also be applied to Windows and Android systems for security at the cloud edge \cite{State-Jour019-1}.

\subsubsection{Time and Safety-Critical Systems}

The challenge of time is of paramount importance in many scenarios \cite{an9}. So many real world safety risks are dependent on the functionality of a computer, such as military bases and ships \cite{Threat-Hunt-Conf001}. The article \cite{Threat-Hunt-Conf001} discusses how these crucial assets are dependent of the success of the security professionals working to protect them. The article further explains that meticulous care needs to be taken to ensure that effective cyber protection measures are taken \cite{Threat-Hunt-Conf001}. With this extreme dependency comes a dire need for strong architecture that works around the physical and logical constraints of each asset \cite{Threat-Hunt-Conf001}. The authors in \cite{Threat-Hunt-Conf001} emphasize the importance that security measures are able to protect large datasets with active cyber detection, and can react in a time efficient manner.

\subsubsection{Software-Defined Networks}
SDN has given the opportunity for improving network resources more efficiently \cite{a12, a11} and providing a foundation for programmability \cite{ Threat-Hunt-Conf007, a10}. SDN is used for complex networks such as the smart cities discussed earlier. The article  \cite{Threat-Hunt-Conf007} explains that SDN works by virtualizing the network and separating the control and data  planes. The authors in  \cite{Threat-Hunt-Conf007} propose that  SDN can be used with ML to enhance threat hunting. This advancement would include intelligent response to DoS, repeat, and man in the middle attacks \cite{Threat-Hunt-Conf007}. The article  \cite{Threat-Hunt-Conf007} concludes that intelligent model will be able to manage the rising amount if traffic demands while maintaining quality of service.

\subsubsection{Critical Infrastructures}

Intrusion Detection and Prevention Systems (IDPD) is studied in \cite{Threat-Hunt-Conf009} and the authors have identified many valuable advantageous solutions, but are also new and have some security gaps. The identified faults include zero-day attacks, unknown anomalies and false positives \cite{Threat-Hunt-Conf009}. The authors in \cite{Threat-Hunt-Conf009} explain that critical infrastructures need supporting mechanisms to fill in the gaps. A web-based platform called TRUSTY is proposed in \cite{Threat-Hunt-Conf009}. TRUSTY is a tool capable of collecting, storing and analyzing the detection results of IDPD for many different industries \cite{Threat-Hunt-Conf009}. The authors in \cite{Threat-Hunt-Conf009} explain that network traffic data is collected by Honeypots that can then be transferred to the supporting tool to provide more insight to the threats. Using a tool like this will fill the vulnerable gaps discussed previously \cite{Threat-Hunt-Conf009}.

\subsection{Enabling Technologies} \label{tech}
\subsubsection{Threat Intelligence}

Crypto-ransomware is an emerging threat that works by a threat accessing the victim’s data and encrypting it \cite{State-Jour021}. The article \cite{State-Jour021} discusses how attacker keep encrypted data and render it useless to the victim unless they pay a ransom for it to be unencrypted. If this threat is detected early it could be stopped before compromising the entire machine’s data \cite{State-Jour021}. The authors of \cite{State-Jour021} go on to explain that timely detection depends on how quickly and accurately a system log can be mined to hunt and stop the attacker \cite{State-Jour021}. In \cite{State-Jour021} using Sequential Pattern Mining was proposed to find Maximal Frequent Patterns will provide useful techniques in ransomware hunting. Stream Data Mining techniques can also be used to cut down the ransomware response time \cite{State-Jour021}. The article \cite{State-Jour021} concludes that the patterns can also help distinguish ransomware families and identification, which can then help develop profiles for threat actor groups.

With the diversity of threats, there is no way to perfectly protect your data. Cyber threat intelligence (CTI) is a technique proposed in \cite{HHH-Jour001} and can be used to quickly map out an overview of the security threat at hand. This method has been widely adapted by many organizations to gain insights and prepare for the cyberattacks that may be used to target them \cite{HHH-Jour001}. The authors of \cite{HHH-Jour001} explain that CTI is done by first designing a threat intelligence meta-schema to visualize the commonalities of infrastructure nodes. The next step is to design a meta-path and graph to measure the similarity between infrastructure nodes and convolutional networks to identify the threat types associated with each node \cite{HHH-Jour001}. It was concluded that this CTI method will provide assistance to analysts with the extensive amount of analysis work they are required to do, which will result in efficient protection from threats \cite{HHH-Jour001}.

\subsubsection{Artificial Intelligence}
Existing IoT systems have monolithic architecture which is implemented in a single solution \cite{State-Jour014}. In \cite{State-Jour014} the authors stated that the architecture is useful for its intended system but has poor scalability, and an overload may crash other functions which would require more work and maintenance. Because of these challenges, the authors of \cite{State-Jour014} proposed a new microservice architecture has been introduced. The new architecture splits up a single solution into manageable components so that they can run and be managed independently of each other \cite{State-Jour014}. The article \cite{State-Jour014} explains that the amount of microservices can be changed depending on the load requirements. The implementor should also take into consideration “service discovery, interservice communication, data integrity, security, monitoring and health check, and quality assurance” \cite{State-Jour014}. The previously mentioned tool TRUSTY developed in \cite{Threat-Hunt-Conf009}, can also be used for tracking and ensuring the integrity of data. 
The authors of \cite{State-Jour014} concluded that AI can also help with the collection and prediction of valuable information generated by devices and humans. AI-enabled microservices can enhance the scaling of data sources and improve intelligent services \cite{State-Jour014}.

\subsubsection{Data Analytics}
Regarding data analytics, we can consider this work \cite{Threat-Hunt-Conf009}. This paper considered a Reinforcement Learning (RL) model, which decides about the number of honeypots that can be deployed in an industrial environment. Indeed, this decision is converted into a Multi-Armed Bandit (MAB), which is solved with the Thompson Sampling (TS) method. The evaluation analysis demonstrates the efficiency of the proposed method.

\subsection{Related Hunting Techniques}\label{related}
\subsubsection{Malware Hunting}

A Remote Access Trojan (RAT) is a malware that is installed on a victim’s machine and gives C2 access back to the attacker \cite{State-Conf112}. The article \cite{State-Conf112} explains that an attacker will be able to  steal sensitive information, spy on the victim and control the hosts machine with the access from these tools. The RATs have been advanced so that firewalls have a very difficult time detecting them and they may be left unnoticed for a period of time \cite{State-Conf112}.  The authors of \cite{State-Conf112} also note that malware detection is a very important part of threat hunting and needs to be updates as the malware types are also advanced. Tools can be used to scan the host machine to look for forensic evidence or artifacts that may have been left behind by the attacker \cite{State-Conf112}. In \cite{State-Conf112} it is explained that running special tools in the memory and on the hard drive can provide an effective solution to finding advanced malware.

\subsubsection{Attack Hunting}

In  \cite{State-Conf113} Ethereum smart contracts are explained ad programs that are used in blockchain and managed by a peer to peer network. The article \cite{State-Conf113} describes that there are attacks that use re-entry that aim to steal the crypto currency Ether, which is stored in deployed smart contracts. In \cite{State-Conf113} a mitigation tactic for this type of threat that includes dynamic analysis that creates its own smart contracts is investigated. Though this approach requires a high cost and thorough preparation so the researchers looked for another method \cite{State-Conf113}.  The solution from \cite{State-Conf113} is RA, a re-entrancy analyzer, which is a combination of symbolic execution and equivalence  checking by a satisfiability modulo theories solver to analyze smart contracts vulnerabilities against the threat concerns. RA excels above other tools as it can perform analysis without prior knowledge of attack patterns and without cost \cite{State-Conf113}. RA can also verify the presence of vulnerabilities without the use of smart contracts and is not expected to produce false positive or false negative results \cite{State-Conf113}.

Another tool for blockchain analysis is described by the authors of \cite{CPBRH-Conf001} is Decentralized finance (DeFi). Defi has become one of the most common application for public blockchains, such as Ethereum discussed previously  \cite{CPBRH-Conf001}. DeFi allows users to freely participate in complex blockchain transactions with the use of contracts and low costs \cite{CPBRH-Conf001}. But due to the flexibility of DeFi there is an inevitable amount of threats introduced. A solution called BLOCKEYE has been introduced in  \cite{CPBRH-Conf001} that mitigates these threats. The article  \cite{CPBRH-Conf001} explains the capabilities of BLOCKEYE, which include an automated security analysis process, which produces a threat reasoning solution for data flow, as well as a transaction monitor that is installed off chain for an at risk DeFi project. Transactions are sent to the project for security analysis and potential threats are flagged when a violation is detected in a critical invariant \cite{CPBRH-Conf001}. The authors of  \cite{CPBRH-Conf001} conclude that BLOCKEYE provides a solution for flagging real-time attacks from end-to-end of the project.

\subsubsection{Vulnerability Hunting}
Researchers are working to find a solution that detects buffer overflow vulnerabilities in C code. One approach proposed in \cite{State-Conf111} utilizes a light weight smart fuzzer to generate string based inputs. This concept was developed on the evolutionary algorithm which is a combination of genetic algorithm and evolutionary strategies \cite{State-Conf111}. The approach from these researchers is an advancement from others as it is able to generate inputs though it does not know the constraints explicitly \cite{State-Conf111}. The smart fuzzer is able to automatically generate inputs while generating inputs dynamically \cite{State-Conf111}. The authors of \cite{State-Conf111} conclude that this model does have some vulnerabilities, specifically to BoF attack, but there are certain checks and protection measures that that can be implemented to mitigate the threats.

\cite{State-Conf111}
An Evolutionary Computing Approach for Hunting Buffer Overflow Vulnerabilities: A Case of Aiming in Dim Light  

\section{Threat Hunting: Challenges}\label{Chall}

With a good understanding of threat hunting, it is now crucial to consider some challenges and how to overcome them for an effective search.subsection 6.1 describes some tools for categorizing threats and the associated indicators, events, and actors. Subsection 6.2 discusses locations of possibly suspicious activity and methods that have been used to protect each.

   \begin{figure*}
    \centering
    \includegraphics[width=15cm]{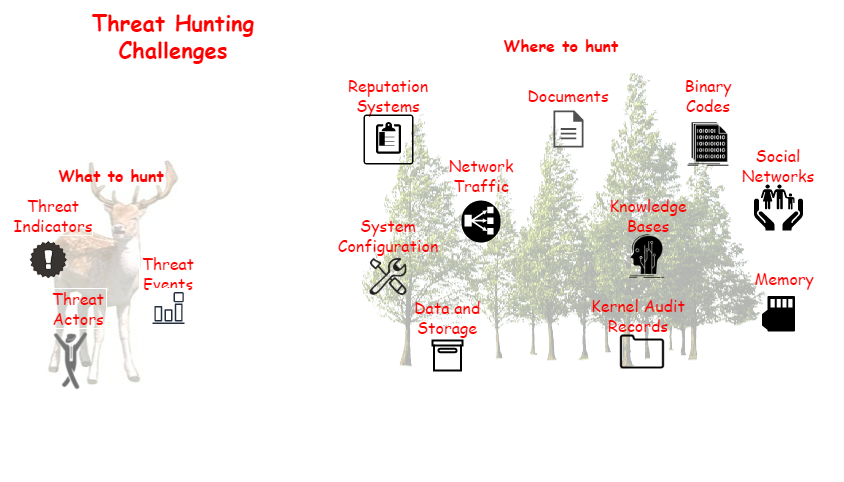}
    \caption{Threat Hunting Challenges}
    \label{fig:challenges}
\end{figure*}

\textcolor{black}{The major challenge of threat hunting is the amount of potential locations that a hunter may find a threat or attack. Threat indicators, events and actors are all important for hunters to be aware of and monitoring. These threats are explored in \ref{what} and specifically include discussion of APTs. After understanding what needs to be hunted the potential locations they may be found can be seen under \emph{Where to hunt} in \ref{fig:challenges}. These locations include threat intelligence knowledge bases, social networks, reputation systems, network traffic, and device specific storage, data, audit records, configurations, documents, and binary codes. Specific details on these locations are discussed in \ref{where} }

\subsection{What to Hunt}\mbox{}\label{what}

\subsubsection{Threat Indicators}

One prevalent challenge in cybersecurity is knowing where to hunt for threats as information is usually hidden and scattered \cite{State-Conf001}. In  \cite{State-Conf001} it is explained that an understanding the threat landscape is crucial to discovering important information about the threat actor. The authors of  \cite{State-Conf001} propose a knowledge graph approach to provide a foundation for building a threat profile. They explain that this graph could recognize a named threat entity and pull related information or related entities from media or blog posts \cite{State-Conf001}. This process would assist in the automation of developing a comprehensive threat landscape and assist researchers in timely decision making \cite{State-Conf001}. The article \cite{State-Conf001} closes with the notion that this model may also give opportunity for future development to build stronger and more accurate automated systems.

Another way to increase threat awareness is through communication. The sharing of CTI has grown in popularity and has proven to be one of the prominent functions to protect users \cite{State-Conf002}. Unfortunately CTI is not always available or guaranteed to be accurate or trustworthy \cite{State-Conf002}. Security professionals have been looking for a solution to estimate the IoCs in fields such as domain names, URLs, and IP addresses \cite{State-Conf002}. A new method has been proposed in  \cite{State-Conf002} to estimate the maliciousness of IoCs more accurately through a graph convolutional network (GCN)-based approach. The authors explain that the GCN approach can be used to estimate the maliciousness of  individual IoC features and their related information \cite{State-Conf002}. This model provides higher accuracy compared to conventional supervised learning models and graph based methods \cite{State-Conf002}.

Another article \cite{State-Conf003} on threat indication describes how certain threats, such as malware, can go undetected as the attacker can disguise the infiltration deep in the system. This is possible because attackers use techniques, such as packers, to avoid malware detection systems \cite{State-Conf003}. The researchers of \cite{State-Conf003} have invested time in malware analysis to try and understand and prevent these attacks, and use techniques such as reverse engineering. They successfully developed an analysis method to create IoCs in the YARA rule format \cite{State-Conf003}.The authors of \cite{State-Conf003} describe how YARA is implemented as a tool for detecting malware using the indicators that were identified during analysis \cite{State-Conf003}. They further explain that the indicators used are the bytecode of common malware samples, emails, and dictionary attacks \cite{State-Conf003}. When the IoC no longer finds other malware located in the directory it can be considered successful \cite{State-Conf003}.

\subsubsection{Threat Events}
There are a plethora of threats that jeopardize the safety of life, belongings, and communities, which influence the way we behave with our assets \cite{State-Conf004}. In \cite{State-Conf004} operational skills are described and are usually in place to ensure that proper procedure is followed in high stress situations. Organizations typically develop the structure on the foundations of structure, information, and leadership \cite{State-Conf004}. Technology has provided those that plan these procedures the tools and automation for their tasks \cite{State-Conf004}. The authors of \cite{State-Conf004} describe that training programs are working to develop a common approach using systematic and logical procedures but emerging threats are constantly changing and needing updated procedures.

Despite research efforts to mitigate distributed threat events there has been little improvement due to the complex associated challenges \cite{State-Jour001}. A new network centric approach proposed in \cite{State-Jour001} uses a holistic approach, consisting of spatially interconnected elements for the detection of these dispersed threats. This new model is compromised of two layer random fields to explain the time variance of the traffic forwarding behavior \cite{State-Jour001}. The authors of  \cite{State-Jour001} explain that the bottom layer describes the connections between the network elements under the action of network elements. The top layer involves each network element’s traffic patterns that are molded by the underlying behaviors \cite{State-Jour001}. The authors of \cite{State-Jour001} conclude that their work is not limited to a specific scenario, so it can be used for an array of different threat detection scenarios.

To recognize the threat events a probabilistic model for intentional behavior has been developed in \cite{State-Conf005} to enable the generation of models used for methods in Technical Intelligence and Operational Intelligence. The components included are measurement, inference, planning and control \cite{State-Conf005}. The authors of  \cite{State-Conf005} explain that it is best to solve the predicting agent components separately then combine them systematically later on to gain the best view of the threat.

Another way to identify threat events is looking back into history. There are many current and historic social events that have triggered cyberattacks, producing a need for a social dimensional threat model \cite{State-Conf006}. To do this, researchers in  \cite{State-Conf006} investigated the likelihood of a cyberattack, which would lead to the ability to form predictions about impending attacks. Then the methods for the attack, the estimated targets, and the duration could be inferred \cite{State-Conf006}. The researchers of \cite{State-Conf006} proposed a concept lattice that can be used to organize the known information and then find patterns, commonalities, and specific details about historical social events that can be applied to current.

A heuristic situation recognition approach has been developed in \cite{State-Conf007} as another event detection system. The purpose of this approach is to enhance Security Management Systems and aims to eliminate the severity of consequences from human faults and sensor faults \cite{State-Conf007}. The authors in \cite{State-Conf007} explained that the approach can be used for a variety of sensor/ alarm dependent systems and can be used while on or offline. The article concludes that model quickens the decision making of the detection systems and also does not require specific AI modelling formalization \cite{State-Conf007}.

\subsubsection{Threat Actors}

Understanding the mass amounts of threat actor profiles is a daunting task to complete effectively. The researchers in \cite{State-Conf008} have developed a knowledge graph of threat actors to assist in organizing all the information. This graph was created by building an ontology of threat actors and the named entity recognition system, which can be used to  automatically extract cybersecurity information from the internet \cite{State-Conf008}. The article explains that the  harvested information is automatically used to create a knowledge graph for the threat actor \cite{State-Conf008}. This knowledge can be used to identity the group when the attack is or has happened.

Cybersecurity professionals are challenged with formulating ways to describe non-uniform and unstructured data, but a solution to this is required to enrich sharing in the field \cite{State-Conf009}. Commonly agreed upon vocabularies for characterizing threat actors is crucial for information exchange at a higher, more meaningful, level \cite{State-Conf009}. The researchers in y agreed upon vocabularies for characterizing threat actors is crucial for information exchange at a higher, more meaningful, level \cite{State-Conf009} have developed a method to automatically infer the types of threat actors based on personas and understand their behavior while accounting for potential changes in the future \cite{State-Conf009}. A set of characterization attributes can enhance the data held about a threat actor \cite{State-Conf009}. The article  \cite{State-Conf009} explains that using deductive reasoning cybersecurity professionals can now automatically infer an attacker’s nature.

\subsection{Advanced Persistent Threats (APTs)}\mbox{}
Although Web Application Firewalls have capabilities to defend against known methods they are still vulnerable to web APTs that use an array of unfamiliar attack methods \cite{APT-Jour002}. To fight against the advanced threats a Web-APT-Detect system was created by the authors of  \cite{APT-Jour002}. The system implements self translation machine through an encoder and decoder using an attention mechanism \cite{APT-Jour002}. They article explains that these mechanisms can increase the quality of the systems used to recognize malicious patterns in HTTP requests \cite{APT-Jour002}. 

Another mechanism POIROT, is a casual correlation aided semantic analysis system that has been developed to unify the current systems created to detect APTs \cite{APT-Jour004}. The authors of  \cite{APT-Jour004} explain that POIROT can detect multi-stage threats over a extended period of time and automatically determines relation between similar events which assists in alert chains. The program also uses Latent Dirichlet Allocation to help reproduce the ATP scenario for further information \cite{APT-Jour004}. The article summarizes that with these tools researchers will be able to map the potential influence of each attack stage \cite{APT-Jour004}.

Data Backup and Recovery (DBAR) techniques when defending against these advanced groups. In  \cite{APT-Jour005} specific techniques have been developed as an ATP defense mechanism which can overcome the drawbacks of previous repair models \cite{APT-Jour005}. Using dynamic model characterization the problems with the previous model can be reduced and made more cost effective \cite{APT-Jour005}. The researchers in  \cite{APT-Jour005} proposed an SDN enabled simulation environment needs to be used for the DBAR strategies to be implemented effectively.

Repair is an additional important factor when considering recovery. But there are many challenges associated with developing effective repair strategies after an attack has been executed. In  \cite{APT-Jour005} it is stated that organizations that do the repairs need to establish an evolution model for the expected state. This is where the impact of the lateral movement can be judged \cite{APT-Jour005}. The authors in  \cite{APT-Jour005} assume that the attacker will try to maximize the potential benefit so the organization will try and minimize loss. Next the organization develops a system for calculating an equilibrium for repair \cite{APT-Jour005}. Once the APT response has been established, game theory can be applied and ‘Nash equilibrium can be determined \cite{APT-Jour008}. The article explains that the organization must finally perform comparisons with random attacks to conclude the equilibrium of APTs \cite{APT-Jour005}. The authors summarize by stating the collected information can be used to address APT response problems and insecurities.

To mitigate the impact of an APT attack organizations must be able to isolate infected devices There is a challenge while doing this that requires a custom dynamic quarantine and recovery plan \cite{APT-Jour007}. In \cite{APT-Jour007} the optimal control theory is explained. The researchers have proposed a concept for normal potential optimal (NPO) control \cite{APT-Jour007}. This concept functions by comparing NPO with the old scheme they found an increase of effectiveness for defending against attacks \cite{APT-Jour007}.

Adversarial Tactics Techniques and Common Knowledge (ATT\&CK) is a matrix that has been developed previously to detect APT attacks that is based on an array of potential classifiers. The classifiers include credential dumping techniques, behavioral features and other systems \cite{APT-Jour009}. The Strange Behavior Inspection (SBI) has been introduced in \cite{APT-Jour009} with the purpose of detecting an attack before it advances to a severe issue. SBI detects the APT at the first target during attack and prevents them from taking full control of the victim’s machine \cite{APT-Jour009}.

\subsection{Where to Hunt} \label{where}
\subsubsection{Threat Intelligence Knowledge Bases}

Open source information sources are a great way to gain intelligence. Open-source Cyber Threat Intelligence (OSCTI) has provided researchers with threat behaviors and have highlighted the gaps that need to be addressed \cite{State-Jour013}. In  \cite{State-Jour013} THREATRAPTOR is developed as a system to automate threat hunting using the OSCTI foundation. THREATRAPTOR’s functionalities include an unsupervised NLP pipeline for unstructured data, a domain specific query language, an automated TBQL query synthesis method, and a query engine for big data \cite{State-Jour013}. The researchers explain that THREATRAPTOR has been tested using a broad set of data from attack cases and has proved to be accurate and effective  \cite{Threat-Hunt-Conf008}.

The authors of  \cite{Threat-Hunt-Conf003} developed the tool CTI ANT. CTI ANT is an automated system that has been developed to analyze Chinese CTI and increase threat intelligence visibility. CTI ANT includes an automatic classification system, a recommendation system, and a web API to label threat techniques \cite{Threat-Hunt-Conf003}. The findings from the system include a cybersecurity article classifier, a cyber topic recommendation system, and a MITRE ATT\&CK detector \cite{Threat-Hunt-Conf003}. Together these tools provide cybersecurity professionals with strong knowledge bases for protection and further research.

\subsubsection{Documents}

A malicious document detector, named Forensor, has been developed in \cite{State-Conf010} to help with protection against attacks. Forensor uses open source tools to inspect file formats and retrieve objects inside then decrypt using simple methods and determines if the contents is malicious \cite{State-Conf010}. Forensor uses a emulator to verify the presence of shellcode, if it is present then the file is malicious \cite{State-Conf010}. This tool can be very useful for determining the legitimacy of documents before opening them and possibly infecting the device.

\subsubsection{Binary Codes}

A solution has been presented in  \cite{State-Conf011}that addresses the challenge of determining IoCs. The new code includes hashes that are organized using an inverted index, which provides constant time for getting the files that contain code hash. The solution also provides support for getting code similarity and quick updates \cite{State-Conf011}.

There is another method proposed in \cite{Threat-Hunt-Conf015} that uses static and dynamic hunting techniques to distinguish malicious and benign binaries quicky. This method can identify signature-based anomalies, and pinpoint the behavior changes that arise when malware is activated \cite{Threat-Hunt-Conf015}. The static hunting is used to classify discovered artifacts based on comparison with other known patterns \cite{Threat-Hunt-Conf015}. Dynamic hunting is used to find behavioral outliers \cite{Threat-Hunt-Conf015}.

\subsubsection{Network Traffic}

IoT devices are being exploited with botnets such as Mirai \cite{State-Conf012}.  The article \cite{State-Conf012} looks at wide network sessions were used with big data analysis to try and gain an understanding about the severity of these botnets. cNetS is a system developed in \cite{State-Conf012} that can scan a system and alert to the presence of a botnet, and providing information on its location and behaviors in the network \cite{State-Conf012}.

\subsubsection{Network Logs}
\subsubsection{Kernel Audit Records}

POIROT is a system developed in  \cite{Threat-Hunt-Conf014} that takes advantage of the correlations between CTI indicators to map the steps of a good campaign. Kernel audits are used as a source for correlation of data flow and to model the hunting with pattern matching \cite{Threat-Hunt-Conf014}. A similarity metric that assess the fir of a query graph and a provenance graph  was used to further the research \cite{Threat-Hunt-Conf014}.  The researchers in \cite{Threat-Hunt-Conf014} discovered that PIRIOT can search the inside of graphs in a few minutes and effectively portray that CTI correlations are substantial artifacts in threat hunting. The PIRIOT algorithm is designed to find the threat behavior of the provenance graph of the kernel audits \cite{Threat-Hunt-Conf014}.

\subsubsection{System Configurations}
Role Based Access Control (RBAC) security is reliant o the quality of the roles and finding the correct roles is often difficult \cite{State-Conf013}. A new spectral clustering algorithm has been proposed in \cite{State-Conf013} to account for user similarities and abnormalities. An abnormal configuration hunting method is then introduced to search for improper assignments and then make suggestions for the proper configuration regarding the clustering results \cite{State-Conf013}. At first a permission sensitive model with adjustable scaling was proposed, but then the researchers in \cite{State-Conf013} decided on an abnormal configuration for hunting rules was implemented.

\subsubsection{Memory}
	
Logs in the memory are an important place to investigate for attacks. Automated security tools formulate logs to organize the patterns used to make new tools \cite{State-Jour002}. The authors of \cite{State-Jour002} explained that some design tools are limited in the ways they collect the logs. To create valuable logs researchers have proposed generating malicious code alerts and binding memory forensic processes for active threat hunting \cite{State-Jour002}. The article \cite{State-Jour002} concludes that these methods will assist in the generation of log memories and only have the malicious entries produce RAM alerts.

\subsubsection{Data and Storage}

SWIFT is a threat investigation system which provides a high traffic causality tracking and real time causal graph generation capabilities \cite{Threat-Hunt-Conf013}. An intelligent memory database was designed in \cite{Threat-Hunt-Conf013} to enable memory savory graph storage and online tracking using as little disk as possible. The researchers conceptualized a storage system that manages forensically relevant parts of the of the graph and disregarding the rest to the disk \cite{Threat-Hunt-Conf013}. The authors in \cite{Threat-Hunt-Conf013} describe how asynchronous cache eviction policy that computes the most untrustworthy section of the causal graph and stores it in the main memory. This tool assists with threat hunting in data storage.
Furthermore, the author of \cite{FGJPA-Book001} provides comprehensive descriptions of how to approach these searches and important features to look for. A comprehensive list of locations and techniques are provided 	\cite{FGJPA-Book001}.

\subsubsection{Reputation Systems}

A new feedback system using Mean Bisector Analysis and Cosine Similarity (MBACS) has been proposed in \cite{State-Conf070} to find malicious users in online reputations. The article \cite{State-Conf070} describes how MBACS is a useful tool for detecting a malicious rating and compile the true ratings, by focusing on rating values and the user domain. \cite{State-Conf070} concludes that using MBACS can reduce the impact of unfair ratings and preserve an accurate reputation.

\subsubsection{Social Network}

A decoy plan has been introduced in \cite{State-Jour018} to assist in mitigating threats against Supervisory Control and Data Acquisition (SCADA). The planned decoy will take unknown threats so that researchers can find the gaps in data \cite{State-Jour018}.Professionals can use SCADA to increase detection abilities more than compared to traditional mechanisms \cite{State-Jour018}. This tool will help defenders identify malicious attacks in social networks.
Another system for automatic verification is XHunter \cite{State-Conf014}. The authors of \cite{State-Conf014} explain that XHunter computes proper conditions for common strings and compares those to ones observed from social networks. The results found many unidentified malicious attacks in web applications \cite{State-Conf014}.

\subsection{Other Challenges}

  Towards other challenges, there is a viable work,  \cite{State-Jour018}, that shows last line of defense in reliability through inducing cyber threat hunting with deception in SCADA networks.

\section{Future Roadmap: The Promise of AI}\label{Fut}

We anticipate that research on threat hunting will move towards quantum-inspired AI-assisted and bio-inspired AI-assisted threat hunting in near future. Our reason for such an anticipation is the existence of the trends discussed in Subsections \ref{sub1} through \ref{sub6}.

\subsection{AI-Assisted Malware Hunting}\label{sub1}
Cybersecurity is a growing problem in today's world due to the use of computers and the Internet by more people. Malware is one of the most prevalent threats on the Internet as stated by antivirus companies. This paper \cite{State-Conf107}  used deep learning algorithms for grouping and creating novel malware samples to tackle this problem \cite{State-Conf107}.

\subsection{AI-Assisted Threat Management}\label{sub2}

Threat management is a challenging and growing task that professionals face. AI has been introduced in a few different ways to help assist in the processing of different aspects \cite{a13}. The article \cite{State-Conf109} explains that an outcome based learning model can take advantage of judgement, decision making, and learning theories to identify the behavior’s of emerging threats. Though the authors explain that the model may need more work to function more effectively with other parameters and factors \cite{State-Conf109}.

One AI based approach is deep learning neural nets, which have been introduced to organize threat intelligence sources quickly and accurately \cite{State-Conf108}. In \cite{State-Conf110} CNNs are used with the Google TensorFlow program to examine images and train a ML model to recognize malicious files \cite{State-Conf110}. The algorithm used in \cite{State-Conf110} can classify the images based on if the user information was malicious or not. 
Similarly, a deep RNN solution has been proposed in \cite{State-Jour028} as a short-term memory that makes use of randomization to limit random network initialization. The final method in \cite{State-Jour028} has reduced complexity and results in better accuracy and higher MCC and AUC compared to previous methods. 
AI can be used to deal with Fifth era (5G) systems. 5G sees a huge amount of data traffic that needs updated security. A new framework has been suggested in  \cite{ABCD-Jour001} and recognizes the dangers regarding 5G and uses learning strategies with facts from the network stream to ensure security. These highlights register all types of traffic and monitor for malicious traffic evidence \cite{ABCD-Jour001}.

There is a dire need for an optimized ML cyberthreat detection model to minimize false positive rates \cite{DCBA-Conf001}. An efficient ML algorithm was developed in \cite{DCBA-Conf001} and functions to gather data, then uses prediction, classification, and forecasting algorithms to produce analytical and empirical evaluations.

There are many different AI systems that can be used in the future. DBN, decision tree, and SVM are the ML techniques that have been used to evaluate some of the most threatening cyberthreats in the cyberspace \cite{DCBA-Conf002}. The article \cite{DCBA-Conf002} describes how these techniques have been used in spam detection, intrusion and malware detection and the precision and accuracy of each has been measured and compared.Having a deep understanding of these tools will allow for better future development.

An advancement on current architecture is proposed in \cite{DSBA-Conf003}. A cascaded CNN architecture with a binary classifier has been proposed for detection and a multi class model for classification of cyber related tweets \cite{DSBA-Conf003}. The results of the classification test were okay, but the model needs further training and testing for robustness \cite{DSBA-Conf003}. The authors in  \cite{DSBA-Conf003} explain that there has also been a classifier model designed to moderate and generate related IT infrastructure\cite{State-Conf081}. The classifier approach considers two approaches: a model for the IT ensemble and one for several parts of the infrastructure \cite{State-Conf081}. The authors conclude by emphasizing that single classification system is preferable other complex systems, and multi layered nets with SVM attributed the best balance between true positives and negatives \cite{State-Conf081}.

Considering where and who cyberattacks are coming from is an important factor. Deep learning models have been implemented to attribute Threat actors based on threat reports obtained from various Threat Intelligence sources\cite{DCBA-Conf004}. The article \cite{DCBA-Conf004} discusses how neural nets that are used to perform the attribution proved to be more accurate than previous techniques and gave better performance.

Looking back on past attacks can give insight to what may happen in the future. In \cite{DCBA-Conf005} ML was been used to recognize some complex examples of threat information dependency on previous knowledge, and what expectations are present.This ML model can be applied to cybersecurity as a mean to predict, identify and advert the complex threats \cite{DCBA-Conf005}

Fuzzy neural nets have been developed in \cite{DCBA-Conf006} to formulate the conception of intuitionistic fuzzy reasoning. The learning algorithms showed that this method can enhance credibility of threat assessment and improve the quality of each assessment with a high level of accuracy \cite{DCBA-Conf006}.

The Internet of Medical Things (IoMT) is on the rise, and threats against certain systems and protocols is closely following \cite{Jour001}. Researchers in \cite{Jour001} have introduced an intrusion detection and prevention system to automatically reduce and mitigate the threats using ML techniques. They explain that this system created reduces the attack surface and helps detect multi layer cyber attack \cite{Jour001}.

The Insider threat detection via Probabilistic pairwise interaction and Heterogeneous Event’s entity embedding (IPH) is a probabilistic model that plots the likelihood of heterogeneous event sequences \cite{OIRU-Conf001}. The model can preserve nonlinear relationships and compute sequence pairwise interactions \cite{OIRU-Conf001}. Tests have shown that the model is effective and is advantageous compared to previous models \cite{OIRU-Conf001}.

Sorting possible threats is an important factor in proper preparation and response. A attribute classification insider threat detection method was created in \cite{OIRU-Conf002} to detect events and extract features with attribute classifiers and an anomaly calculator for a end to end framework. The researchers conducted tests and the results support this proposition and the performance was validated \cite{OIRU-Conf002}.

A network based model that uses ML methods to profile mobile threats and analyze the network flows for malware connections was proposed in \cite{OIRU-Conf003}. The researchers found that the model can be used to combine outputs and it was found efficient for detecting known and unknown threats \cite{OIRU-Conf003}.

To better the security and stability of the industrial Internet, researchers in \cite{OIRU-Conf004} have investigates the industrial control network traffic threat identification based on ML and uses heuristic methods for selecting parameters and speed up real time detection. They found that these methods are faster and perform better than other identification methods \cite{OIRU-Conf004}.

Large scale companies must ensure that they can manage possible threat scenarios. National Critical Infrastructure has a duty to protect their sensitive information. In \cite{OIRU-Conf005} it is explained that NCI security can be increased using a significant proof of concept system to detect the threats via fitness evaluation by EEG signals. This is done by using deep learning algorithms to classify the range of mental states into the four categories of the risk matrix \cite{OIRU-Conf005}. 
In similar research \cite{OIRU-Conf006}, the tree structure method has been used to analyze user behavior, form feature sequences, and combine the Copula Based Outlier Detection (COPOD). These methods can be used to visualize the difference between notable sequences and outstanding users. The article summarizes by saying that the systems were able to analyze data without limitation on the number of dimensions and had fast computation speeds with little parameters \cite{OIRU-Conf006}.

AI has also been used in threat prevention and sensing engines for two factors that form the critical points \cite{OIRU-Conf007}. The authors of  \cite{OIRU-Conf007} explain that the two relevant factors are intelligent packet inspection and intelligent first reaction. Though the article concludes that no single approach will create a holistic solution for threat prevention and many more are required \cite{OIRU-Conf007}.

A new framework has been proposed in \cite{OIRU-Conf008} for constructing a user-centered ML based insider threat system for many granularity levels. The results of testing the framework showed that the ML based detection system learns from the limited knowledge and can detect unrecognized malicious threats with a high level of accuracy \cite{OIRU-Conf008}.

AI can also be used for network intrusion detection. The article \cite{OIRU-Conf009} describes how network intrusion detection has been advanced with the integration of CNNs using LeNet-5 to classify the threats to the network. This CNN detection method has improved the accuracy when detecting intrusions and used enhanced features to classify threats\cite{OIRU-Conf009}.

Universal communication is important for sharing techniques and advice. Deep cross-lingual models have been used to jointly learn the common representation from two languages \cite{OIRU-Conf010}. The model from  \cite{OIRU-Conf010} exceeds the functionality of previous monolingual models previously used to translate non-English cyberthreats. This tool can be used for corporations to implement universal protocols.  
The authors of \cite{UYRE-Jour002} discuss standardized response protocols. In \cite{UYRE-Jour002} it was hypothesized that a  challenge response protocol provides bettered security on a public domain. The researchers use eight classifiers to demonstrate that the new method has a slight impact on security standards but increases usability and comfort, as well as enhances the advantages compared to current standards \cite{UYRE-Jour002}.

Having a good understanding of data can ensure its accuracy. Unified conceptual and computational framework with progressive learning algorithms can be used for research, analysis and comparison for learning capacity and prediction accuracy for datasets and the cloud domain \cite{UYRE-Jour001}. In \cite{UYRE-Jour001} it is explained that extensive amounts of metrics have been used for predicting the future of domains for security and imaging. The results of \cite{UYRE-Jour001} show a structured framework for automatically generating network threat detection with emerging threats through development.

ML models using decision trees, Bayesian network, and deep learning can be used for quick response and organization of APT attacks on specific datasets \cite{APT-Jour001}. The article \cite{APT-Jour001} emphasized how it is important to consider sensitivity, specificity, accuracy, false negative rate, and F-measure and investigated during the choice and use of each model.

\subsection{AI-Assisted Threat Intelligence}\label{sub3}
The authors of \cite{MZHU-Jour001} explains that the IoT systems require a strong connection between Space, Air, Ground, and Sea networks to suggestion automated services to users and companies. Security and safety issues can arise with these networks if IoT systems are not protected successfully \cite{MZHU-Jour001}. Security experts are now using Threat Intelligence to comprehend cyberattacks and to protect SAGS networks with AI design. This study \cite{MZHU-Jour001} offers a novel TI structure constructed on deep learning that can identify cyberthreats within SAGS networks \cite{MZHU-Jour001}.

Increasingly, our nation's critical infrastructure is being attacked by cyberattacks\cite{JSTU-Conf001}. This study \cite{JSTU-Conf001} offers, develops, and examines a Cyberthreat Intelligence structure. Results of simulated attacks on a dataset from an Industrial Control System displays along with the extracted indications of compromise \cite{JSTU-Conf001}.

\subsection{AI-Assisted Threat Hunting}\label{sub4}
As SDN has gained popularity, it has introduced a tendency of novel technologies in the networking area \cite{Threat-Hunt-Conf007}. In a network environment, SDN provides elasticity and compatibility via splitting the control plane from the data plane by means of virtualizing the network hardware \cite{Threat-Hunt-Conf007}. This report \cite{Threat-Hunt-Conf007} presents a model for advanced threat hunting that merges SDN infrastructure-based threat hunting techniques with ML models for managing network threats including DOS, and MITM attacks\cite{Threat-Hunt-Conf007}.

Because ICPs are complex, large-scale, and varied, identifying cyberthreats is a difficult task\cite{State-Jour016}. The study of \cite{State-Jour016} offers a new federated deep learning model that captures the temporal and spatial features of network data in order to hunt cyberthreats against ICPs. This paper presents a descriptive micro-service placement method to improve micro-service utilization by leveraging the collaborators' computational resources to address the latency problem of an ICSP \cite{State-Jour016}.

The purpose of \cite{CMLF-Conf001} work is to introduce a secure self-optimizing, self-adapting system-on-chip (S4oC) architecture design and optimization structure. By making real-time modifications, we can reduce the impact of attacks to the smallest amount including hardware Trojans and side channels \cite{CMLF-Conf001}. S4oC is vulnerable to many security measures and attacks since it learns to reconfigure itself. In addition, the target applications' data types and patterns, environmental settings, and sources of variation are combined \cite{CMLF-Conf001}.

Digital society and Internet continue to be at risk of malware \cite{State-Jour022}. Currently, malware hunting methods usually rely on one solo view including using dynamic information or op-codes only \cite{State-Jour022}. In order to overcome these restrictions,\cite{State-Jour022} offer a multi-view learning approach that uses op-codes, byte-codes, header information, permission, attacker's intent, and API calls to search for malicious programs. Authors of \cite{State-Jour022} showed that their method is very precise with low false positive rates with experiments conducted on several Windows, Android, and IoT platforms 
\cite{State-Jour022}.

Network intrusion detection research based on ML has a huge problem because the experimental environments do not reproduce real-world scenarios where unidentified attacks are continuously emerging. Since they have used one data set for training and testing, the discovery influence is overestimated because all test attack types are identified in training, while the test cases will be alike to the training data. The paper presents a novel method to create test data with updated traffic with attacks types not found in training data
\cite{VCXZ-Jour001}.

\subsection{Quantum-Inspired AI}\label{sub5}

In this subsection we discuss studies that are inspired by quantum including Reinforcement Learning (RL) regarding UAV-Mounted Wireless Networks and Robot Navigation, Neural Network for Data Classification and  Multi-directional Associative Memory.

 The objective of this article \cite{Quant-AI-Jour001} is to examine a wireless communication with the satellite transmission situation in which a vehicle that has no crew and fly on air functions as a base station to gather data from users on the ground. By using quantum-inspired RL the direction planning issue is improved without previous knowledge of the ground users including their locations, channel state information, and transmission power \cite{Quant-AI-Jour001}. 
Here is another example of Quantum-Inspired reinforcement learning (QiRL) \cite{Quant-AI-Jour002} article offers a new training model caused by quantum computation for profound RL with experience repetition. To reach an equilibrium between investigation and exploitation, the offered quantum-inspired experience replay system selects experiences from the replay buffer adaptively and according to the complication of the experience as well as the number of times it has been replayed \cite{Quant-AI-Jour002}. Moreover, for navigation control of autonomous mobile robots, an original quantum-inspired RL (QiRL) algorithm is suggested by \cite{Quant-AI-Jour004}. A probable action selection strategy and a novel reinforcement approach in QiRL are encouraged, by quantum measurement failure and quantum computation domain strengthening. A number of simulated experiments of Markovian state transition confirmed that QiRL is stronger than out-of-date RL when compared to learning degrees and early conditions\cite{Quant-AI-Jour004}. In \cite{Quant-AI-Jour003}, research in quantum-inspired computing has considerably enhanced the potentials of traditional algorithms. Generally, quantum information processing in neural frameworks is represented by quantum-inspired Hopfield associative memory. A quantum inspired multidirectional associative memory (QMAM) with a single report learning model, and QMAM with a self-convergent repetitive learning model (IQMAM) is presented in  \cite{Quant-AI-Jour003}.

It was discussed in \cite{Quant-AI-Jour005} that neural networks (NN) execute based on a diversity of factors including the structure, early weight, quantity of concealed layer neurons, and learning proportion. A challenging problem is enhancing NN grouping performance without altering its structure\cite{Quant-AI-Jour005}. In terms of precision, correctness, and uniqueness, the offered Q-FNN model\cite{Quant-AI-Jour005} outperforms state-of-the-art approaches on 15 genuine standard datasets.

\subsection{Bio-Inspired AI}\label{sub6}

The human system is an amazing tool. Implementing biological system processes into computing systems may have many advantages \cite{a14}. In \cite{Fut-Bio-Jour001} Visual attention prediction (VAP) was explained as an important challenge for computer vision. A new approach to VAP is proposed in \cite{Fut-Bio-Jour001} that combines low-level features and high-level semantics similar to a human eye for visual mapping. The article \cite{Fut-Bio-Jour001} explains that the new VAP method performs stronger than the other current methods.

Moving from the eye, we now look at the face. Facial aesthetics has peaked the interests of researchers \cite{Fut-Bio-Jour002}. The previous standard was not able to accurately represent human perception \cite{Fut-Bio-Jour002}. The authors of \cite{Fut-Bio-Jour002} designed a biological based project to trace eye movements and recognize human features. The system in \cite{Fut-Bio-Jour002} uses this data to create a Bio-Inspired Facial Aesthetic Ontology, and involve a CNN to train a set of human feature detectors. The system can then accurately categorize if a face is considered beautiful, and list the determining reasons \cite{Fut-Bio-Jour002}. The new model is able to identify very specific parts of the face which provides extra support for the decision.

Beings other than humans can also provide interesting techniques for computers to replicate. In \cite{Fut-Bio-Jour003} hummingbirds are studied as they have very interesting movement patterns. The article \cite{Fut-Bio-Jour003} describes how hummingbird movement can be used to develop RL. RL can assist or even take over conventional stabilization techniques \cite{Fut-Bio-Jour003}. The robot using RL from \cite{Fut-Bio-Jour003} was able to use rapid escape maneuvers and complete full body flips.

RL has also been used in \cite{Fut-Bio-Jour005} for autonomous navigation where the system learns to interact with the environment and learn behaviour for maximum benefit. The researchers in \cite{Fut-Bio-Jour005} created a system with very few rewards so the RL algorithm is highly trained in identifying the goal reward. This training adds additional robustness for the prediction methods in the CNN model \cite{Fut-Bio-Jour005}.

The article \cite{Fut-Bio-Jour004} explores bio-inspired methods to analyze email datasets. Multiple ML models including Naive Bayes, SVM, Random Forest,Decision Trees, and Multi-layer Perceptron were considered in \cite{Fut-Bio-Jour004} to evaluate these datasets. Though these tools are powerful, the bio-inspired algorithms Particle Swarm Optimization and the Genetic algorithm proved better for this analysis \cite{Fut-Bio-Jour004}.

Relating back to the topic of IoT, the authors of \cite{Fut-Bio-Jour006} implemented human interaction techniques for bio-inspired self-learning coevolutionary algorithm (BSCA). BSCA essentially just optimizes the interactions between the internet connected devices by reducing the energy used, increase diversity of intersections, and search methods to cope with multiple requests \cite{Fut-Bio-Jour006}. The study \cite{Fut-Bio-Jour006} proved that BSCA performs better than the current algorithms for high dimensional problems \cite{Fut-Bio-Jour006}.

Learning practices for machines and humans are both important areas that need constant development. BOLE is a MATLAB interactive learning environment that was developed in \cite{Fut-Bio-Jour007} for the development of automated aerial vehicle path planning. BOLE helps with learning as it focuses on fundamental concepts and breaks down problems into introduction, recognition, practice and collaboration based on the problems complexity \cite{Fut-Bio-Jour007}. The article \cite{Fut-Bio-Jour007} concluded that BOLE is a excellent tool to compliment traditional teaching and because it is bio-inspired it is very intuitive for human use.

\begin{figure}
    \centering
    \includegraphics[width=10cm]{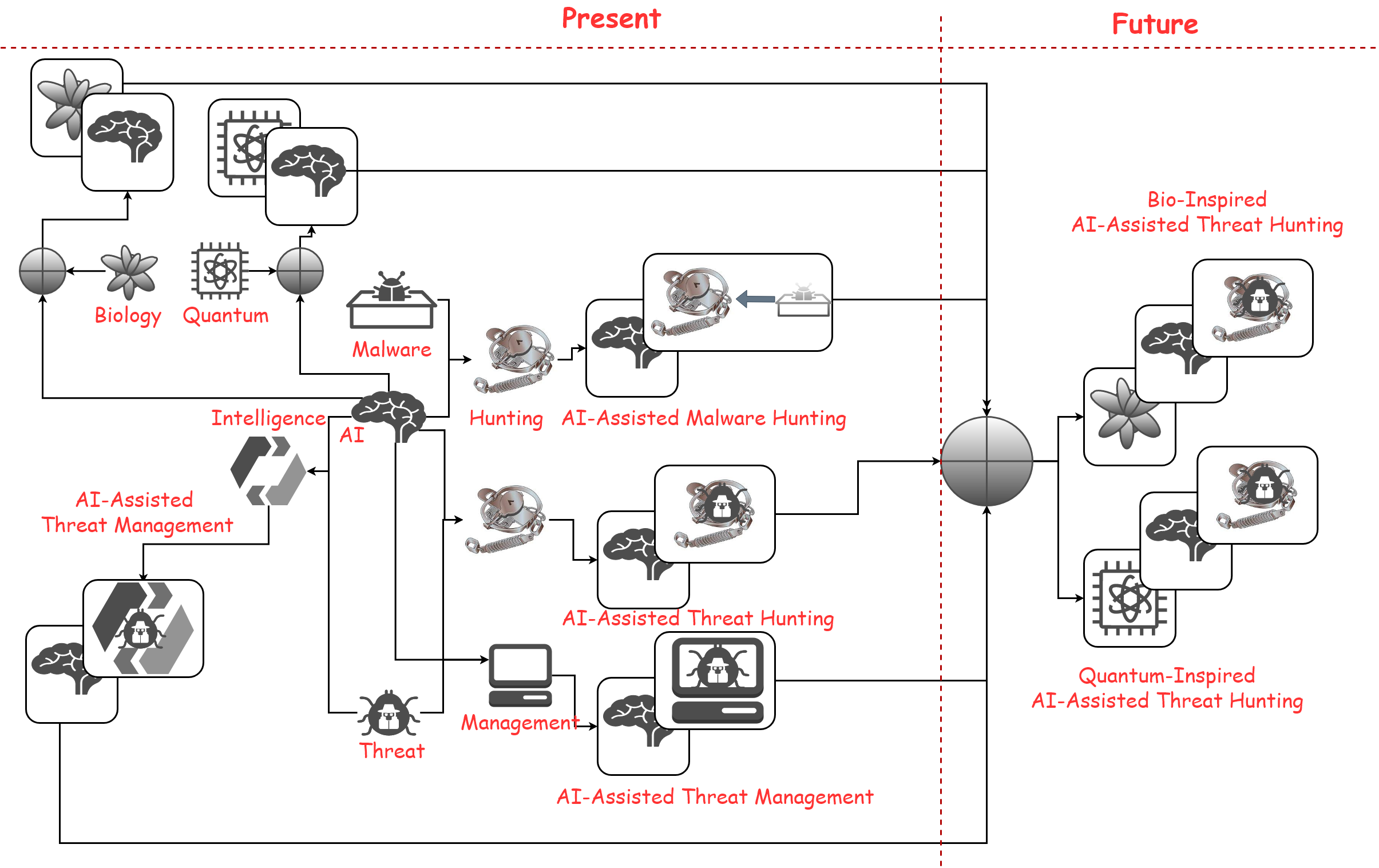}
    \caption{The Future Roadmap of Threat Hunting}
    \label{fig:sec7.6}
\end{figure}

The topics seen throughout the treat hunting life cycle are all generally moving towards AI based models which take inspiration from biological specimens and quantum processes. The development from current procedures to the future can be seen in Figure \ref{fig:sec7.6}. The three main areas of current threat hunting is the hunting of malware, threat management, and threat intelligence. Combining the features of each of these we can see the development to the future of threat hunting.

Figure \ref{fig:sec7.6} shows how the complex topics discussed in this article correlate with each other and influence the future of threat hunting.

\textcolor{black}{The present topics of threat hunting can be advanced using bio-inspired and quantum-inspired AI assistance (Figure \ref{fig:sec7.6}). \emph{AI-assisted malware hunting} (Section \ref{sub1}), \emph{AI-assisted threat management} (Section \ref{sub2}), \emph{AI-assisted threat intelligence} (Section \ref{sub3}), \emph{AI-assisted threat hunting} (Section \ref{sub4}) are all present topics that are leading to the integration of \emph{quantum-inspired AI} (Section \ref{sub5}), and \emph{bio-inspired AI} (Section \ref{sub6}) to better threat hunting practices. }

\section{Conclusion} \label{Conc}

\textcolor{black}{This paper provides a comprehensive review of the current threat landscape, specifically focusing threat hunting through an array of specified methods and tools.
The life cycle stages and ecosystem are thoroughly discussed with support of recent research in those areas.
We also identify many challenges that are currently seen in present literature regarding threat hunting.
The review shows that the future of threat hunting will utilize bio-inspired AI and quantum inspired AI. The AI techniques have great potential to broaden the scope and increase automation of threat hunting to assist the cybersecurity professionals.
This work provides researchers with a holistic understating of present work and gives insight as to the future direction of the field. }

\bibliographystyle{IEEEtran}
\bibliography{References}

\end{document}